\documentclass[10pt,draftclsnofoot,onecolumn]{IEEEtran}

\pdfoutput=1
\usepackage{float}
\usepackage{amsfonts}
\usepackage{amsmath}
\usepackage{amsthm}
\usepackage{amssymb,latexsym}
\newtheorem{definition}{Definition} 
\newtheorem{lem}{Lemma}

\newtheorem{thm}{Theorem}
\newtheorem{prob}{Problem}
\newtheorem{remark}{Remark}
\newtheorem{assum}{Assumption}
\usepackage{graphicx}
\usepackage{float}
\usepackage{subfig}
\usepackage{caption}
\usepackage{mathrsfs}
\usepackage{xcolor}
\usepackage{cite}

\usepackage{booktabs}
\newcommand{\ra}[1]{\renewcommand{\arraystretch}{#1}}

\newcommand{\sB}{\mathcal{B}}

\newcommand{\sP}{\mathcal{P}}
\newcommand{\sQ}{\mathcal{Q}}

\newcommand{\sS}{\mathcal{S}}

\newcommand{\sU}{\mathcal{U}}
\newcommand{\sV}{\mathcal{V}}
\newcommand{\sW}{\mathcal{W}}
\newcommand{\sX}{\mathcal{X}}
\newcommand{\sY}{\mathcal{Y}}

\newenvironment{proof-sketch}{\emph{ Sketch of Proof:}}{\small\qed\\}

\begin{document}

\title{Approximate Safety Verification and Control of Partially Observable Stochastic Hybrid Systems}
\author{Kendra Lesser, \IEEEmembership{Member, IEEE}, Meeko Oishi, \IEEEmembership{Member, IEEE}
\thanks{K. Lesser is with the Department of Computer Science, University of Oxford, OX1 3QD, United Kingdom (e-mail: kendra.lesser@cs.ox.ac.uk).}
\thanks{M. Oishi is with the Department of Electrical and Computer Engineering, University of New Mexico, Albuquerque, NM, USA (e-mail: oishi@unm.edu).}
\thanks{This material is based upon work supported by the National Science Foundation under Grant Number CMMI-1254990, CNS-1329878, and CMMI-1335038 and by the Air Force Office of Scientific Research through a Summer Faculty Fellowship.  Any opinions, findings, and conclusions or recommendations expressed in this material are those of the authors and do not necessarily reflect the views of the National Science Foundation.}}
\maketitle

\begin{abstract}

Assuring safety in discrete time stochastic hybrid systems is particularly difficult when only noisy or incomplete observations of the state are available.  We first review a formulation of the probabilistic safety problem under noisy hybrid observations  as a dynamic program over an equivalent information state.  Two methods for approximately solving the dynamic program are presented.  The first method approximates the hybrid system as an equivalent finite state  Markov decision process, so that the information state is a probability mass function.  The second approach approximates an indicator function over the safe region using radial basis functions, to represent the information state as a Gaussian mixture.  In both cases, we discretize the hybrid observation process and generate a sampled set of information states, then use  point-based value iteration to under-approximate the safety probability and synthesize a suboptimal control policy.  We obtain error bounds and convergence results in both cases, assuming switched affine dynamics and additive Gaussian noise on the continuous states and observations.  We compare the performance of the finite state and Gaussian mixture approaches on a simple numerical example.

\end{abstract}

\section{Introduction}

Safety critical systems, such as aircraft, satellites, and electricity grids, often rely on sensors to measure their state and their environment.   The true state of the system may not be measurable, or may be corrupted by noise, as quantified by a so-called observation process.  The controller must choose actions based only on the information contained in the observation process.  The nature of safety critical systems dictates a need for formal methods to accurately  assess the system's ability to meet rigorous safety requirements, and also to synthesize controllers that guarantee performance to a desired level (correct by design).  It is therefore paramount that the controller exploit information from the observation process, to obtain theoretical safety guarantees that are as  accurate as possible.

Reachability analysis, which determines whether a system's state remains within a given safe region and/or reaches a target set within some time horizon, has been used extensively as a tool for verification and controller synthesis for hybrid systems \cite{oishi1}, \cite{Prand1}, \cite{Mitchell1} and extended to stochastic hybrid systems (SHS) 
\cite{Abate1}, \cite{summers}.
There has been little focus, however, on reachability analysis for partially observable SHS.  
While there has been some work on deterministic hybrid systems with hidden modes \cite{Verma2012} or uncertain systems with imperfect information on a partial order \cite{Ghaemi2013}, reachability analysis of a partially observable SHS has been approached only theoretically \cite{Ding2013}, \cite{Lesser2014}. 
This, along with our previous work \cite{Lesser2015_HSCC} provides the first computational results for both controller synthesis and verification of safety specifications for partially observable SHS.

Existing computational results for reachability analysis of fully observable SHS are also limited. The safety problem for a discrete time SHS (DTSHS), which considers only whether the state of the system can be controlled to remain within a safe region of the state space, can be formulated as a multiplicative cost stochastic optimal control problem \cite{Abate1}, and solved in the same manner as a Markov decision process (MDP).  Unfortunately, solutions via dynamic programming  \cite{bertsekas2} require evaluation of a value function over all possible states, which is infinite when those states are continuous.   Discretization procedures can be employed to impose a finite number of states, as in \cite{Abate10} and \cite{SoudjaniSiam13}, which present rigorous uniform and  adaptive gridding methods for verification of DTSHS.  Similarly, approximate abstractions of the original stochastic model to an equivalent system that has the same properties are presented in \cite{Julius09}, \cite{Franzle11}, and \cite{Zamani14}. 
Even so, current applications are limited to those with only a few discrete and continuous states.

The safety problem for a partially observable DTSHS (PODTSHS) can similarly be formulated as a partially observable MDP (POMDP).  However, POMDPs are plagued by dimensionality on an even greater scale than MDPs.  The common approach to solving POMDPs is to replace the growing history of observations and actions by a sufficient statistic, often called the belief state, which, for a POMDP with an additive cost function, is the distribution of the current state conditioned on all past observations and actions  \cite{bertsekas2}.  This belief state is treated as the perfectly observed true state, and MDP solution methods can then be applied.  However, with a continuous state space, the belief state is a function defined over an infinite domain, and it is impossible to enumerate over all such functions.  Therefore the study of efficient, approximate solutions to POMDPs is essential.

Although finding the solution to a general POMDP is hard \cite{Lusena2001}, many algorithms for approximating solutions to finite state POMDPs have been developed.  These mainly rely on point-based value iteration (PBVI) schemes that only consider a subset of the belief space to update the value function used in the dynamic program (for a survey of PBVI algorithms, see \cite{shani13}).  Because the value function is piecewise-linear and convex \cite{sondik} (and so equivalently represented by a finite set of vectors), sampling from the belief state provides a systematic way of storing a finite subset of those vectors.   Such methods must be tailored to continuous state POMDPs because of the dimensionality of the belief state.  

Other than discretizing the state space and solving an equivalent finite state POMDP, many existing methods for continuous state POMDPs assume the belief state is Gaussian (e.g. \cite{Brooks06}, \cite{Zhou2010}), and represent the belief state in a parameterized form which is then discretized and solved as a discrete state MDP.  When the belief state cannot adequately be represented using a single Gaussian, a Gaussian mixture may be used instead.  An equivalent point-based algorithm for continuous-state POMDPs using Gaussian mixtures is presented in \cite{Porta06}, and demonstrated on a stochastic hybrid system with hidden modes in \cite{Brunskill2010}.  

The safety problem for a PODTSHS is further complicated because 
the belief state is not the conditional distribution of the current state of the system \cite{Ding2013}, \cite{Lesser2014}, but must also include the distribution of a binary variable that indicates whether the state of the system has remained within a safe region up to the previous time step.  This, coupled with the stochastic hybrid system dynamics, makes accurately representing the belief state as a single Gaussian impossible.  

We formulate the safety problem for a PODTSHS as a POMDP, and investigate representations of the belief state in either vector or Gaussian mixture form through  finite- and continuous-state approximations to the PODTSHS.   These representations allow us to exploit point-based methods developed for POMDPs.

This paper extends our previous work \cite{Lesser2015_HSCC} in several ways.  First, we validate the use of POMDP solution techniques for  reachability analysis of a PODTSHS, by demonstrating that the value function is convex and admits a function representation related to the piecewise-linear vector representation of a finite state POMDP.  Second, we present a finite state approximation to the DTSHS (presented in \cite{Lesser2015_HSCC} without proofs) that allows the belief state to take vector form under certain conditions, and show convergence for the approximation.  Third, we preserve the continuity in the hybrid state space through a  Gaussian mixture representation for the belief state, and approximate the indicator function that represents the safe region using Gaussian radial basis functions.  In this case, we provide an error bound as a function of the integrated error (1-norm in the function space $L^1$) of the indicator function approximation.  Our solution method  converges to the true solution from below, using either the finite or continuity-preserving belief state.  We demonstrate both approaches on a temperature regulation problem.

The rest of the paper is organized as follows.  Section \ref{sec:background} 
relates the safety problem for a PODTSHS to optimal control of a POMDP.  Section \ref{sec:abstraction} justifies the use of POMDP solution techniques, and presents the finite and Gaussian mixture approximations to the safety problem for a PODTSHS (as well as error bounds).  Section \ref{sec:pbvi} describes the use of point-based approximation techniques, through sampling of belief states and discretization of the observations.  We present a numerical example in Section \ref{sec:example}, and concluding remarks and directions for future work in Section \ref{sec:conc}.  All proofs can be found in the Appendix.

\section{Background}\label{sec:background}

\subsection{Notation and Probability Overview}

A probability space $(\Omega, \mathcal{F},\mathbb{P})$ consists of a sample space $\Omega$, a $\sigma$-algebra $\mathcal{F}$ defined over $\Omega$, and a probability measure $\mathbb{P}$ that assigns probabilities to events in $\mathcal{F}$.  For $\Omega = \mathbb{R}^n$, we presume $\mathcal{F} = \mathcal{B}(\mathbb{R}^n)$, the Borel $\sigma$-algebra on $\mathbb{R}^n$.  The probability measure $\mathbb{P}$ maps elements $B\in\mathcal{B}(\mathbb{R}^n)$ to the interval $[0,1]$.  The density $p$ associated to $\mathbb{P}$ is defined according to the Lebesgue measure as $\mathbb{P}(B) = \int_B p(x)\,dx$.  

We denote expected value by $\mathbb{E}$.  A probability measure or expected value induced by a control policy $\pi$ (to be defined later), is $\mathbb{P}^{\pi}$ or $\mathbb{E}^{\pi}$, respectively.  For a space $\mathcal{X}$, $\mathcal{X}^n = \mathcal{X}\times\dotsc\times\mathcal{X}$ is the $n$-times product space of $\mathcal{X}$.  The state of a system at time $n$, $x_n$, takes values in state space $\mathcal{X}$.  We use $|\cdot|$ to denote absolute value for $x\in\mathbb{R}$ and cardinality for a set $\Omega$.  A vector or function norm is denoted with $\|\cdot\|$.  Unless otherwise specified, $\|\cdot\| = \|\cdot\|_2$.  A random variable $x\sim \mathcal{N}(\mu, \sP)$ follows a Gaussian distribution with mean $\mu$ and covariance $\sP$, and $\phi(x; \mu, \sP)$ represents a Gaussian density with mean $\mu$ and covariance $\sP$ evaluated at $x$.  Finally, $\langle \cdot, \,\cdot\rangle$ denotes an inner product,  with $\langle u, v\rangle = u^Tv$ for vectors $u,v\in\mathbb{R}^n$, and $\langle f, g\rangle = \int f(x)g(x)\,dx$ or $\langle f,g\rangle = \sum_q \int f(x,q)g(x,q)\,dx$ for appropriately defined functions $f$, $g$.

\subsection{Computing Optimal Control Policies for POMDPs}\label{sec:back_pomdp}

The main results of this paper rely on framing a PODTSHS as a POMDP with a hybrid state space, and drawing upon existing results for finite state POMDPs with additive cost objectives.    We therefore present an overview of POMDPs and efficient approximation techniques for their optimal control, and then express a PODTSHS with a safety objective as a POMDP with a multiplicative cost function.






\begin{definition}\label{pomdpDef}(POMDP $\mathcal{J}$) A POMDP is a tuple $\mathcal{J} = (\mathcal{S}, \mathcal{U}, \mathcal{Y}, T, Y, R)$ where
	
	\begin{enumerate}
		\item
		$\mathcal{S}$ is a finite set of states
		\item
		$\mathcal{U}$ is a finite set of possible control inputs 
		\item
		$\mathcal{Y}$ is a finite set of observations
		\item
		$T: \mathcal{S}\times\mathcal{S}\times\mathcal{U}\rightarrow [0,1]$ is a state transition function that assigns a probability measure to state $s_{n+1}$ given state $s_n$ and  control $u_n$ for all $n$: $T(s_{n+1}| s_n, u_n)$
		\item
		$Y: \mathcal{Y}\times\mathcal{S}\times\mathcal{U}\rightarrow [0,1]$ is an observation function that assigns a probability measure to observation $y_n$ given state $s_n$ and control $u_{n-1}$ for all $n$: $Y(y_n| s_n, u_{n-1})$
		\item
		$R: \mathcal{S}\rightarrow [0,1]$ is an initial probability measure over the state space $\mathcal{S}$: $R(s)$
	\end{enumerate}
	
\end{definition}

The state evolves stochastically and is Markovian (e.g., the state at the next time step depends only on the current state and action).  
The information available to the controller at time $n$ is  $i_n = (u_0,\dotsc,u_{n-1},y_1,\dotsc,y_n) \in \mathcal{I}_n = \mathcal{U}^n \times \mathcal{Y}^n$; that is, the controller cannot directly observe the state.
The control input at each time step is selected according to a control policy $\pi$, which maps the available information at each time $n$ onto $\mathcal U$.
\begin{definition}\label{def:Policy}
	For a POMDP $\mathcal{J}$, a  policy $\pi$ for some time horizon $N$ is a sequence of functions, $\pi = \left(\pi_0,\dotsc,\pi_{N-1}\right)$, such that $\pi_n: \mathcal{I}_n\rightarrow \mathcal{U}$.  
\end{definition}
We consider non-randomized policies, i.e. ones that assign a single control input to each possible $i_n$, which  are sufficient for the problem we consider \cite{bertsekas}.  
A control policy $\pi$ induces a probability space $(\Omega, \sigma(\Omega), \mathbb{P}^{\pi})$ over the POMDP with state space $\Omega = \mathcal{S}^n\times \mathcal{Y}^n$, $\sigma$-algebra $\sigma(\Omega)$ on $\Omega$, and probability measure $\mathbb{P}^{\pi}$ based on $T$, $Y$, $R$, and $\pi$.

The execution of a POMDP is as follows.  At time $n=0$, state $s_0$ is produced from initial distribution $R: s\sim R(\cdot)$.  At each subsequent time $n>0$, an observation $y_n$ is produced according to $y_n\sim Y(\cdot|s_n,u_{n-1})$, and added to the list of past observations and control inputs to produce $i_n = (u_0,\dotsc,$ $u_{n-1},y_1,\dotsc,y_n)$.  The control input is chosen according to $u_n=\pi_n(i_n)$, and cost $C(s_n,u_n)$ is accrued.  The next state $s_{n+1}$ is then generated according to $s_{n+1}\sim T(\cdot|s_n,u_n)$.  

The goal is to minimize the expected sum of costs accrued according to function $C$ over a time horizon $N$ by optimally choosing control actions according to the policy $\pi$.  
\begin{equation}\label{sumCost}
\max_{\pi} \mathbb{E}^{\pi}\left[\sum_{n=0}^NC(s_n, \pi_n(i_n))\right]
\end{equation}

Equation \eqref{sumCost} can be solved using dynamic programming, much like for a Markov decision process \cite{bertsekas2}.  The value function, $V_n(i_n)$, represents the expected sum of costs accrued from time $n$ to $N$ given that $i_n$ has been recorded thus far, and is computed recursively backwards in time.  However, since the size of vector $i_n$ increases with $n$, and is difficult to store, the optimal control input and value function can instead be expressed as a function of a belief state.
The belief state is a \emph{sufficient statistic} for the information vector $i_n$ because it condenses all information necessary for making optimal decisions without sacrificing optimality. 
For an additive cost function $C$, the belief state is a function that describes the probability of being in state $s$ given all past observations and actions, $b(s_n) = \mathbb{P}[s_n|i_n]$ \cite{bertsekas2}.  By treating the belief state as the true state of the system, \eqref{sumCost} can be equivalently solved as an  MDP over the belief state.  The optimal policy $\pi^*:\mathfrak{B}\rightarrow \mathcal{U}$ is hence defined in terms of the belief state, with $\mathfrak{B}$ the space of all beliefs.

The optimal policy can be found using a value function over the belief space,
\begin{equation}
V_n^*(b) = \max_{u\in\mathcal{U}}\left \{\sum_s C(s,u)b(s) 
+\sum_y V_{n+1}^*\left(M_{y,u}b\right)\mathbb{P}(y| b,u)\right\}.
\end{equation}
The transition operator $M_{y,u}b$ provides the next belief state $b_{n+1}$ given the current observation, action, and belief state,
\begin{equation}\label{eq:Moperator}
\left(M_{y,u}b\right)(s') = \frac{Y(y|s',u)\sum_{s\in\mathcal{S}}T(s'|s,u)b(s)}{\mathbb{P}(y|b,u)},
\end{equation}
and the likelihood of the observation is $\mathbb{P}(y|b,u) = \sum_{s\in \mathcal{S}}b(s)\sum_{s'\in\mathcal{S}}T(s'|s,u)Y(y|s',u)$.
Sondik \cite{sondik} first showed that for a finite horizon $N < \infty$, the value function at each time $n$ is piecewise linear and convex in the belief state, and thus can be expressed as
\begin{equation}\label{Valpha}
V_n^*(b) =\max_{\alpha_n\in\Gamma_n}\, \sum_s \alpha_n(s) b(s).
\end{equation}

The functions $\alpha_n\in\mathbb{R}^{|\sS|}$, or ``$\alpha$-vectors'', 
characterize the value of being in state $s\in\sS$ at time $n$ when a specific action $u$ is taken, plus the expected sum of future rewards assuming all subsequent actions are chosen optimally.  Because each $\alpha$-vector is associated with a specific action, selection of the optimal $\alpha$-vector in \eqref{Valpha} defines the optimal policy for belief $b$ at time $n$.  The set $\Gamma_n$ of $\alpha$-vectors needed to exactly represent the value function $V_n$ at time $n$  is finite, but grows exponentially, 
since computing $\Gamma_n$ requires calculation of $|\mathcal{U}||\Gamma_{n+1}|^{|\mathcal{Y}|}$ $\alpha$-vectors. An exact $\alpha$-vector representation is therefore often infeasible, and approximate solutions are required.  

We draw upon methods in point-based value iteration (PBVI) \cite{pineau06}, \cite{shani13} because they provide a lower bound to the value function, which is key for our safety verification problem.  
In the most general PBVI method, a finite subset $B\subset\mathfrak{B}$ is selected, then one $\alpha$-vector is generated for each belief point $b^i\in B$, $B = (b^0, b^1,\dotsc, b^m)$, so that $\tilde{\Gamma}_n = (\alpha_n^0, \alpha_n^1, \dotsc, \alpha_n^m)$.  The value at some $b$ not necessarily in $B$ can be approximated by $V_n^*(b) \approx \max_{\alpha_n\in\tilde{\Gamma}_n}\, \sum_s \alpha_n(s) b(s)$ (as compared to \eqref{Valpha}) with $\tilde{\Gamma}_n\subset \Gamma_n$, since we presume that for any $b$ in a neighborhood of $b^j$ the same action will likely be optimal \cite{pineau06}.  Hence $\tilde{\Gamma}_n$ can be generated recursively from $\tilde{\Gamma}_{n+1}$ without enumeration over all possible combinations of observations and subsequent $\alpha$-vectors in $\tilde{\Gamma}_{n+1}$, by using the backup operation

\begin{equation}\label{backup}
backup(b) = \arg\max_{\alpha_n\in\tilde{\Gamma}_n} \sum_{s\in\mathcal{S}} \alpha_n(s)b(s)
\end{equation}
for each $b\in B$.  Essentially, the PBVI algorithm consists of selecting a set of belief points $B\subset \mathfrak{B}$, and repeatedly applying \eqref{backup} to each element of $B$.  For $N$ finite, the backup operator is applied to $B$ $N$ times.

For PBVI methods that have been extended to POMDPs with a continuous state space and discrete observation and action spaces \cite{Porta06}, the $\alpha$-vectors are replaced by $\alpha$-functions defined over the continuous space $S$.  Since observations and actions are finite, the number of $\alpha$-functions remains finite, and the value function is piecewise-linear and convex under the $\alpha$-function representation.  


\subsection{PODTSHS Modeled as a POMDP}\label{sec:back_model}

A PODTSHS has discrete and continuous states with interacting dynamics that are characterized by stochastic kernels, and observations that may also be hybrid.  It can be expressed within the POMDP framework by extending Definition \ref{pomdpDef}.

\begin{definition}
	\label{podtshs}(PODTSHS $\mathcal{H}$). A PODTSHS is a class of POMDP, i.e. a tuple $\mathcal{H}=(\mathcal{S},\mathcal{U}, \mathcal{Y}, T, Y, R)$, in which
	\begin{enumerate}
		\item
		$\mathcal{S} = \mathbb{R}^m\times\mathcal{Q}$ is a hybrid state space with $\mathcal{Q} = \{q_1, q_2,\dotsc,q_{N_q}\}$ a finite set of states with cardinality $N_q$ and continuous state dimension $m$
		\item
		$\mathcal{U}$ is a finite set of control inputs affecting both discrete and continuous state transitions
		\item
		$\mathcal{Y} =  \mathbb{R}^l\times\mathcal{Y}^q$ is a hybrid observation space with $\mathcal{Y}^q \subseteq \mathcal{Q}$ and continuous observation dimension $l\leq m$
		\item
		$T: \mathcal{B}(\mathbb{R}^m)\times \mathcal{Q}\times\mathcal{S}\times\mathcal{U}\rightarrow [0,1]$ is a  stochastic transition kernel that assigns a probability measure over $\mathcal{B}(\mathbb{R}^m)\times\mathcal{Q}$ at time $n+1$ given $s_n$ and $u_n$ for all $n$: $T(\beta,q|s_n,u_n)$ with $\beta\in\mathcal{B}(\mathbb{R}^m)$
		\item
		$Y: \mathcal{B}(\mathbb{R}^{l})\times\mathcal{Y}^q\times \mathcal{S}\rightarrow [0,1]$ is an observation function that assigns a probability measure over $\mathcal{B}(\mathbb{R}^{p})\times\mathcal{Y}^q$ at time $n$ given $s_n$ for all $n$: $Y(\beta,y^q|s_n)$
		\item
		$R: \mathcal{B}(\mathbb{R}^{m})\times\mathcal{Q}\rightarrow [0,1]$ is an initial probability measure over $\mathcal{B}(\mathbb{R}^{m})\times\mathcal{Q}$: $R(\beta,q)$ 
	\end{enumerate} 
\end{definition} 

\begin{remark}
While we presume $\sU$ is finite, a continuous or hybrid input set can be approximated as a finite set
when computing safety probabilities and the optimal policy.
\end{remark}
The state transition kernel comprises a discrete component $T_q$ that governs mode updates, and a kernel $T_x$ for continuous state transitions.  For modeling purposes, we choose to order \eqref{eq:T} such that the discrete mode $q_n$ updates first at each time step, and the subsequent mode $q_{n+1}$ influences the evolution of $x_n$ to $x_{n+1}$. 
\begin{equation}\label{eq:T}
T(\beta, q| s_n,u_n) = T_x(\beta|q_{n+1}, x_n,u_n)T_q(q_{n+1}|s_n,u_n)
\end{equation}
The functions $Y$ and $R$ are also separated into discrete and continuous components,
\begin{align}
Y(\beta,y^q|s_n) &= Y_x(\beta|s_n)Y_q(y^q|s_n), \label{eq:Y}\\
R(\beta,q) &= R_x(\beta)R_q(q).\label{eq:R}
\end{align}
Functions $T_x$, $Y_x$, and $R_x$ are Borel-measurable stochastic transition kernels over $\mathcal{B}(\mathbb{R}^{(\cdot)})$, and $T_q$, $Y_q$, and $R_q$ are standard probability distributions over finite state spaces.
%

We consider specifically a switched affine system, such that the continuous state $x$ evolves according to 
\begin{equation}\label{stateUpdate}
x_{n+1} = A(q_{n+1})x_n + g(q_{n+1},u_n) + v_n.
\end{equation}
The $v_n$ are independent and identically distributed Gaussian random variables for all $n$, $v_n \sim \mathcal{N}(0, \mathcal{V})$.  The matrix $A$ and function $g$ change according to the mode $q_{n+1}$. 
 We assume the discrete observations $y^q$ depend only on $q$, and the observations $y^x$ depend linearly on $x$, corrupted by additive Gaussian noise $w_n$, with $w_n$ independent and identically distributed for all $n$, $w_n\sim\mathcal{N}(0, \mathcal{W})$. 
\begin{equation}
\label{eq:yUpdate}
y_n^x = C(q_{n})x_n + w_n
\end{equation}

Kernels $T_x$, and $Y_x$ admit Gaussian densities 
$\tau_x(x|q_{n+1},x_{n},u_n) = \phi(x; A(q_{n+1})x_n + g(q_{n+1},u_n), \mathcal{V})$, and $\gamma_x(y^x|s_n) = \phi(y^x; C(q_n)x_n, \mathcal{W})$. We also assume $R_x$ admits a Gaussian density $\rho_x(x) = \phi(x; \mu_0, \sP_0)$. For ease of notation, we let $\rho(s) = \rho_x(x)R_q(q)$, $\gamma(y|s) = \gamma_x(y^x|s)Y_q(y^q|q)$, and $\tau(s'|s,u) =$ $\tau_x(x'|q',x,u)T_q(q'|s,u)$ for $s=(x,q)$, $y=(y^x,y^q)$, and $s'=(x',q')$.

%
%


We require that the following Lipschitz properties hold, which are guaranteed for $\gamma_x$, $\tau_x$, and $\rho_x$, given that they are Gaussian densities and $\sU$ is finite.
\begin{equation}\label{eq:lipschitz}
\begin{gathered}
|\tau_x(x'|q',x,u) - \tau_x(\overline{x}'|q',x,u) |\leq h_x^{(1)}\| x' - \overline{x}' \|\\
|\tau_x(x'|q',x,u) - \tau_x(x'|q',\overline{x},u)| \leq h_x^{(2)}\| x - \overline{x}\|\\
|\gamma_{x}(y|x,q) - \gamma_{x}(\overline{y}|x,q)| \leq h_y^{(1)}\| y - \overline{y}\|\\
|\gamma_{x}(y|x,q) - \gamma_{x}(y|\overline{x},q)| \leq h_y^{(2)}\| x - \overline{x} \|\\
|T_q(q'|q,x,u) - T_q(q'|q,\overline{x},u)|\leq h_q\|x - \overline{x}\|
\end{gathered}
\end{equation}


We define the maximum values 
$\phi_v^* = (2\pi)^{-\frac{m}{2}}|\mathcal{V}|^{-\frac{1}{2}}$ and $\phi_w^* = (2\pi)^{-\frac{l}{2}}|\mathcal{W}|^{-\frac{1}{2}}$.

\begin{remark}
	While we impose assumptions of linearity and additive Gaussian noise in \eqref{stateUpdate} and \eqref{eq:yUpdate} to facilitate subsequent derivations, these assumptions can be relaxed in certain cases, which will be highlighted where appropriate. 
\end{remark}

  
\subsection{Safety Problem}\label{sec:back_invariance}

We use stochastic optimal control to find both a control policy that maximizes the probability of the state remaining within a safe region of the state space, as well as an estimate of that probability \cite{Abate1}.  For a compact Borel set $K\subseteq \mathcal{B}(\sS)$, terminal time $N$, and predefined policy $\pi$, the objective to optimize is $p_{\mathrm{safe}}^N(\pi, \rho; K) = \mathbb{P}^{\pi}[s_n \in K \, \forall \,n=0,\dotsc,N|\rho]$.

This objective can be expressed more commonly as an expected value of a multiplicative cost function, 
since for a random variable $X$ and event $A$, $\mathbb{P}[X \in A] = \mathbb{E}[{\bf 1}_A(X)]$, with the indicator function ${\bf 1}_A(X)=1$ if $X \in A$ and ${\bf1}_A(X)=0$ otherwise, as shown in  \cite{Abate1}:
\begin{equation} \label{ERA}
p_{\mathrm{safe}}^N(\pi, \rho; K) ={\mathbb E}^{\pi} \left[\left. \prod_{n=0}^{N}{\bf 1}_K(s_n) \right| \rho\right].
\end{equation}

The maximal safety probability and  optimal policy $\pi^*$ are given by
\begin{align}
p_{\mathrm{safe}}^N(\rho; K) &= \sup_{\pi\in\Pi}p_{\mathrm{safe}}^N(\pi, \rho; K), \label{eq:optViab} \\
\pi^* &= \arg\sup_{\pi\in\Pi} \left\{p_{\mathrm{safe}}^N(\pi, \rho; K)\right\}. \label{eq:optPol}
\end{align}
In the fully observable case, \cite{Abate1} gives a dynamic programming formulation for optimizing \eqref{ERA}, which returns both the maximal safety probability and optimal policy.   We would like to take a similar approach to find both \eqref{eq:optViab} and \eqref{eq:optPol}.  Formally, we would like to solve the following problem.


\begin{prob}\label{probState}
For a PODTSHS $\mathcal{H} = (\mathcal{S},\mathcal{U}, \mathcal{Y}, T, Y, R)$ with a safe set $K\in\sB(\sS)$ and time horizon $N<\infty$, we wish to
\begin{enumerate}
\item
Compute the maximal probability \eqref{eq:optViab} of remaining within $K$ for $N$ time steps.
\item
Compute the optimal policy $\pi^*$ given by \eqref{eq:optPol}.
\end{enumerate}
If the maximal probability and optimal policy cannot be computed exactly (which is quite likely \cite{Lusena2001}), an approximation that produces a suboptimal policy and lower bound on the maximal safety probability is desired.
\end{prob}

 \section{Reformulation using a POMDP}\label{sec:abstraction}

We exploit the PBVI method to solve Problem \ref{probState}, by transforming Problem \ref{probState} into an optimal control problem for a POMDP.  Hence we first show the safety problem for $\mathcal{H}$ can be reduced to a dynamic program, despite a non-standard belief state.  We then show that the $\alpha$-functions and belief states can be approximately represented in closed form and that finite collections of each may be generated and used to approximate \eqref{eq:optViab}, similar to a point-based POMDP solver.

We present two approximations of Problem \ref{probState} for the PODTSHS $\mathcal{H}$: The first discretizes $\mathcal{S}$ to produce a finite state POMDP, and the second preserves continuity in $\mathcal{S}$ by using a Gaussian mixture approach, thus characterizing the PODTSHS by a collection of weights, means, and covariances.

\subsection{Validity of $\alpha$-Function Representation}\label{sec:abs_just}

The multiplicative nature of the cost function for the safety problem \eqref{ERA} renders the belief state for an additive cost POMDP inapplicable, and we derived a different sufficient statistic $\eta = (\eta_0,\dotsc,\eta_N)$ for Problem \ref{probState} in \cite{Lesser2014}.  This sufficient statistic produces a modified conditional distribution of the current state that includes the probability that all past states are in the safe set $K$.
\begin{equation}\label{eq:sigma}
 \eta_n(\rho, i_n)=  \mathbb{P}^{\pi}[x_n\in \beta, q_n = q, s_0,\dotsc,s_{n-1}\in K\mid \rho, i_n]
\end{equation}

We define the \emph{information state} as the function $\sigma_n(x_n,q_n)\in \Sigma_n\subseteq L^1$ (where $L^1$ is the space of integrable functions) associated with the probability distribution produced by $\eta_n(\rho, i_n)$, so that $\int_{\beta}\sigma_n(z,q)\,dz = \eta_n(\rho,i_n)$ for all $q\in\mathcal{Q}$, $\beta\in\mathcal{B}(\mathbb{R}^{m})$. Note that the information state is distinct from the belief state (e.g.  the conditional distribution of the current state).  The information state updates recursively with a bounded linear operator $\Phi$ (for proof see \cite{Lesser2014})
\begin{equation}\label{sigma_rec}
\begin{aligned}
\sigma_0 &= \rho \\
\sigma_n &= \Phi_{y_n,u_{n-1}}\sigma_{n-1}
\end{aligned}
\end{equation}
where $\Phi_{y,u}\sigma$ is given by
\begin{equation} \label{Phi}
\left(\Phi_{y,u}\sigma\right)(s') = \frac{1}{p(y|\sigma, u)}\gamma(y|s',u)\int_K \tau(s'|s,u) \sigma(s)\,ds .
\end{equation}
In comparing \eqref{eq:Moperator} to  \eqref{Phi}, the latter integrates over the compact hybrid set $K$, as opposed to a summation over finite set $\mathcal{S}$.

We define a dynamic programming recursion over $\sigma$ as
\begin{equation}\label{optDP}
\begin{aligned}
V_N^*(\sigma_N) &= \langle \sigma_N, {\bf 1}_K \rangle \\
V_n^*(\sigma_n) &= \max_{u\in\mathcal{U}} \mathbb{E}^{\pi}\left[ V_{n+1}^*(\Phi_{y,u}\sigma_n)\right]
\end{aligned}
\end{equation}
with solution $V_0^*(\rho) = p_{\mathrm{safe}}^N(\rho; K)$.  The optimal policy is $\pi^* = (\pi_0^*,\dotsc,\pi_{N-1}^*)$, with $\pi_n^*(\sigma_n) = \arg\max_{u\in\mathcal{U}}V_n^*(\sigma_n)$
for all $n\in [0,N]$.

\begin{lem}\label{lem:sigAlpha}
	For any $n$, the value function \eqref{optDP} can be written as
	\begin{equation*}
		V_n^*(\sigma_n) = \sup_{\alpha_n\in\Gamma_n}\, \langle \alpha_n, \sigma_n \rangle.
	\end{equation*}
\end{lem}

Lemma \ref{lem:sigAlpha} leads to the following representation of the value function evaluated at some $\sigma\in\Sigma$.
\begin{gather}
	V_n^*(\sigma) = \max_{u\in\mathcal{U}}\int_{\mathcal{Y}}\langle \alpha_{y,u,\sigma}, \sigma_n\rangle \,dy,\label{eq:Vint}\\
	\alpha_{y,u,\sigma}(s) =\int_{\mathcal{S}}\alpha_{n+1}^{*(y)}(s')\gamma(y|s')\tau(s'|s,u){\bf 1}_K(s)\,ds'\label{eq:alpha_yus}
	\end{gather}
	with $*(y)$ used to denote the observation-dependent optimal $\alpha$-function in $\Gamma_{n+1}$ that maximizes $\langle  \alpha_{n+1}, \Phi_{y,u}\sigma_n \rangle$ (the value function $V_{n+1}(\sigma_{n+1})$)  for a specific observation $y$.  The set of $\alpha$-functions at time $n$ is 
	\begin{equation}\label{Gamma_n}
	\Gamma_n = \bigcup_{\sigma\in\Sigma}\left\{\int_{\mathcal{Y}}\alpha_{y,u^*,\sigma}\,dy\right\}
	\end{equation}
	with $u^*$ the control inputs chosen according to \eqref{eq:Vint}.

\begin{lem}\label{lem:sigConvex}
The value function \eqref{optDP} is convex in $\sigma$ for all $n=0,\dotsc,N$, $\sigma_n^1$, $\sigma_n^2$ $\in L^1$ and $0\leq\lambda\leq1$:
\begin{equation*}
V_n^*(\lambda\sigma_n^1 + (1-\lambda)\sigma_n^2)\leq \lambda V_n^*(\sigma_n^1) + (1-\lambda)V_n^*(\sigma_n^2).
\end{equation*}
\end{lem}
Lemmas \ref{lem:sigAlpha} and \ref{lem:sigConvex} show that the value function \eqref{optDP} is convex and admits an $\alpha$-function representation, hence $\mathcal{H}$ is amenable to POMDP solution techniques.  However, we cannot use Lemma \ref{lem:sigAlpha} to solve Problem \ref{probState} directly, since $\Gamma_n$ is not finite and the $\alpha$-functions and information states have no common structure.


\subsection{Finite State Approximation}\label{sec:abs_finite}

We first consider a finite state POMDP \cite{Abate10}, whose solution converges to the true safety probability \eqref{eq:optViab} and optimal policy \eqref{eq:optPol}. The state space $\mathcal{S}$ is discretized to obtain a vector representation of $\alpha$ and $\sigma$. The observation space is unchanged (i.e. hybrid), because the set of observations only affects the finiteness of sets $\Gamma_n$ and $\Sigma_n$.  We defer discussion of producing finite collections of $\Gamma_n$ and $\Sigma_n$ to section \ref{sec:pbvi}.

Given safe set $K \in \mathcal{B}(\mathcal{S})$, let $K = \bigcup_{q\in\mathcal{Q}} K_q \times \{q\}$.  Denote $\lambda = \max_{q\in\mathcal{Q}}\mathcal{L}(K_q)$, the finite Lebesgue measure of $K_q\subset\mathbb{R}^{m}$.  Each $K_q$ is partitioned into a finite number of subsets, so that $K_q = \bigcup_{i=1}^{m_q}K_{i,q}$, with $K_{i,q}$ pairwise disjoint (i.e. $K_{i,q}\cap K_{j,q} = \emptyset$ for all $i\neq j$).  Finally, let $\delta_{i,q}^x$ be the diameter of partition $K_{i,q}$ so that $\delta_{i,q}^x = \sup\{\| x - \overline{x} \| :  x, \overline{x}\in K_{i,q}\}$, with $\delta^x = \max_{i,q}\delta_{i,q}^x$.

The partition of $K$ is denoted by $\mathcal{G}^s = \bigcup_{i=1,\dotsc,m_q, q\in\mathcal{Q}}\mathcal{G}_{i,q}^s$ with $\mathcal{G}_{i,q}^s = K_{i,q}\times  \{q\}$.  Each element $\mathcal{G}_{i,q}^s$ has a representative point $( x^{i,q}, q)$ and the set $K_{\delta} = \{(x^{i,q}, q) :  i = 1,\dotsc, m_q, q \in \mathcal{Q}\}$ is the discrete representation of $K$.  We do not consider here how the points $(x^{i,q},q)$ are selected, but an example is provided in Section \ref{sec:example}.  The function $\xi: K \rightarrow K_{\delta}$ maps a state $s\in\mathcal{G}_{i,q}^s$ to its  representative point $(x^{i,q},q)$ and the function $\Xi: K_{\delta} \rightarrow K$ is the set-valued map from discrete point $(x^{i,q},q)$ to its corresponding set $\mathcal{G}_{i,q}^s$.  We will abuse notation slightly and interchangeably write $\xi(s) = (x^{i,q},q)$ and $\xi(x) = x^{i,q}$, since $q$ is mapped to itself, and hence the property of interest is the mapping from $x$ to $x^{i,q}$.  The discrete state space is defined as $Z_{\delta} = K_{\delta} \bigcup \{\psi_s\}$, with $\psi_s$ a single variable that represents all states $s \in \mathcal{S}\backslash K$.

\begin{definition}\label{POMDPapprox}(POMDP approximation to PODTSHS, $\hat{\mathcal{H}}$). The POMDP approximation is a tuple  $\hat{\mathcal{H}} = (Z_{\delta}, \mathcal{Y}, \mathcal{U}, \tau_{\delta} , Y, \rho_{\delta})$ where
\begin{enumerate}
\item
$Z_{\delta}$ is a finite set of discrete states
\item
$\mathcal{Y}$ is as defined in Definition 3
\item 
$\mathcal{U}$ is as defined in Definition 3
\item
$\tau_{\delta}: Z_{\delta} \times \mathcal{U} \times Z_{\delta}   \rightarrow [0,1]$ is a discrete state transition function that assigns probabilities to elements of $Z_{\delta}$
\item
$Y: \mathcal{B}(\mathbb{R}^{l}\times \mathcal{Y}^q\times Z_{\delta}\rightarrow [0,1]$ is as defined in Definition 3, but is conditioned only on values $z \in Z_{\delta}$ rather than $s\in\mathcal{S}$
\item
$\rho_{\delta}: Z_{\delta}\rightarrow [0,1]$ is a function that assigns probabilities to elements of $Z_{\delta}$ at time zero
\end{enumerate}
\end{definition}

We define the transition function as
\begin{equation}\label{discreteSKer}
\tau_{\delta}(z'| z, u) = \begin{cases}
T(\Xi(z')|z,u),&\text{if $z'\in K_{\delta}$ and $z\in K_{\delta}$} \\
1 - \sum_{\overline{z}\in K_{\delta}} T(\Xi(\overline{z})|z, u), &\text{if $z' = \psi_s$, $z\in K_{\delta}$}\\
1, &\text{if $z' = \psi_s$ and $z = \psi_s$ } \\
0,&\text{if $z' \in K_{\delta}$ and $z = \psi_s$ }
\end{cases}
\end{equation}
with $\sum_{z'\in Z_{\delta}}\tau_{\delta}(z'|z,u)=1$, and the initial distribution $\rho_{\delta}$ on $Z_{\delta}$ as
\begin{equation}\label{discreteRho}
\rho_{\delta}(z) =  \begin{cases}
R(\Xi(z)), &\text{ if $z \in K_{\delta}$ }\\
1 - \sum_{\overline{z} \in K_{\delta}} R(\Xi(\overline{z})) &\text { if $z = \psi_s$ }
\end{cases}
\end{equation}

The discrete probability space is $(\Omega_{\delta}, \sigma(\Omega_{\delta}), \mathbb{P}_{\delta}^{\pi_{\delta}})$ with $\Omega_{\delta} = Z_{\delta}^{N+1}\times \mathcal{Y}^{N}$, $\sigma(\Omega_{\delta})$ the $\sigma$-algebra on $\Omega_{\delta}$, and $\mathbb{P}_{\delta}^{\pi_{\delta}}$ the probability measure uniquely defined by $\rho_{\delta}$, $Y$, $\tau_{\delta}$, and a control policy $\pi_{\delta} = (\pi_0^{\delta},\dotsc, \pi_{N-1}^{\delta})$, $\pi_n^{\delta} : \Sigma_{n,\delta}\rightarrow \mathcal{U}$, with $\Sigma_{n,\delta}$ the set of all information states $\sigma_{n,\delta}$ defined on $Z_{\delta}$ at time $n$.

We further define the operator $\Phi^{\delta}_{y,u}$ and the intermediate vector $\alpha^{\delta}_{y,u,\sigma_{\delta}}$ for any $y\in\mathcal{Y}$, $u\in\mathcal{U}$, $z',z \in Z_{\delta}$ as
\begin{gather}
\left({\Phi}_{y,u}^{\delta}\sigma_{\delta}\right)(z') = \frac{1}{p(y|\sigma_{\delta},u)}\gamma(y|z')\sum_{z\in K_{\delta}}\tau_{\delta}(z'|z,u)\sigma_{\delta}(z) \label{eq:phiDisc} \\
\alpha^{\delta}_{y,u,\sigma_{\delta}}(z) = \sum_{z'\in K_{\delta}}\alpha_{n+1,\delta}^{*(y)}(z')\gamma(y|z')\tau_{\delta}(z'|z,u){\bf 1}_{K_{\delta}}(z)\label{eq:alphDisc}.
\end{gather}
The safety problem for $\hat{\mathcal{H}}$ is to find $p_{\mathrm{safe},\delta}^N(\rho_{\delta}; K_{\delta}) = \sup_{\pi_{\delta}\in\Pi_{\delta}} \mathbb{P}^{\pi_{\delta}}\left[ z_n \in K_{\delta}, \, \forall\, n=0,\dotsc,N 
|\rho_{\delta}\right]$, which is solved by formulating the information state $\sigma_{n,\delta}$ and the value function $V_{n,\delta}^*: \sigma_{n,\delta}\rightarrow [0,1]$ for $n=0,\dotsc,N$. 

The discrete information state represents a probability mass function over $Z_{\delta}$, and can be expressed as an integral over an equivalent density (just as $\tau_{\delta}(z'|z,u) = T(\Xi(z')|z,u)$).
\begin{equation}\label{eq:sigInt}
\sigma_{n,\delta}(z) = \begin{cases}
\int_{\Xi(z)}\hat{\sigma}_n(s)\,ds, \mbox{ if $z \in K_{\delta}$ } \\
\int_{\mathcal{S}\backslash K} \hat{\sigma}_n(s)\,ds, \mbox{ if $z = \psi_s$ }
\end{cases}
\end{equation}
with $\hat{\sigma}_n(s)$ given by
\begin{equation}\label{eq:sigHat}
\hat{\sigma}_n(s') =\begin{cases}
\rho(s'), &\text{if $n = 0$ } \\
\left(\hat{\Phi}_{y,u}\hat{\sigma}_{n-1}\right)(s') =  \frac{1}{p(y|\hat{\sigma}_{n-1},u)}\gamma(y|\xi(s')) \int_K \tau(s'|\xi(s),u)\hat{\sigma}_{n-1}(s)\,ds, &\text{if $n>0$}
\end{cases}
\end{equation}
This can be verified by substituting the expression for $\tau_{\delta}$ in terms of $\tau$ into \eqref{eq:phiDisc} and using an induction argument.

The value function is
\begin{equation} \label{eq:valDiscrete}
\begin{aligned}
V_{N,\delta}^*(\sigma_{N,\delta}) &= \sum_{z\in K_{\delta}} \sigma_{N,\delta}(z)\\
V_{n,\delta}^*(\sigma_{n,\delta}) &= \max_{u\in\mathcal{U}}\int_{\mathcal{Y}}V_{n+1,\delta}^*(\Phi_{y,u}^{\delta}\sigma_{n,\delta})\mathbb{P}_{\delta}(dy|\sigma_{n,\delta},u)
\end{aligned}.
\end{equation}
The maximum probability of remaining within $K_{\delta}$ over $N$ time steps is
\begin{equation}\label{eq:discreteViab}
p_{\mathrm{safe}, \delta}^N(\rho_{\delta}; K_{\delta}) = V_{0,\delta}^*(\rho_{\delta}).
\end{equation}

The safety probability for the finite state approximation $\hat{\mathcal{H}}$ converges to the true solution as grid size parameter $\delta^x$ tends to zero.  To show this, we first describe the error between the continuous information state $\sigma_n$ and the vector approximation $\sigma_{n,\delta}$.   



\begin{thm}\label{thm:sigDiscrete}
The density $\hat{\sigma}$ defined in \eqref{eq:sigHat} satisfies
\begin{equation*}
|\sigma_n(s) - \hat{\sigma}_n(s)| \leq \eta_n^{\sigma}\delta^x
\end{equation*}
for all $s\in\mathcal{S}$, $\sigma_n\in\Sigma_n$, and $\eta_n^{\sigma}$ given by
\begin{equation*}
\eta_n^{\sigma} = \sum_{i=0}^{n-1} c_{1,i}\left(\prod_{j=i+1}^{n-1} c_{2,j}\right),
\end{equation*}
with $c_{1,i} = \min\{\frac{1}{p(y|\sigma_i,u)}, \frac{1}{p(y|\hat{\sigma}_i, u)}\}[\phi_v^*h_y^{(2)} + \phi_w^* h_x^{(2)} + \phi_w^*\phi_v^*h_q]$, $c_{2,j} = \min\{\frac{1}{p(y|\sigma_j,u)}, \frac{1}{p(y|\hat{\sigma}_j, u)}\}\phi_w^*N_q\lambda$.
\end{thm}


To prove convergence of the value function $V_{n,\delta}^*$ to $V_n^*$, we must first show that integration over the infinite spaces $\sY$ and $\sS$ results in a bounded solution. 

Consider the following two  lemmas regarding integration of $\gamma_x$ and $\tau_x$ over unbounded sets $\mathbb{R}^l$ and $\mathbb{R}^m$, respectively.
\begin{lem}\label{lem:finiteIntY}
For any $x, \overline{x}\in K_{i,q}$, for all $i=1,\dotsc,m_q$, $q\in\mathcal{Q}$, the following holds:
\begin{equation*}
\int_{\mathbb{R}^{l}} |\gamma_x(y^x|x,q) - \gamma_x(y^x|\overline{x},q)|\,dy^x \leq \left[\beta_{1,i,q}^yh_y^{(2)} + \beta_{2,q}^y\right] \delta_{i,q}^x
\end{equation*}
with $\beta_{1,i,q}^y = \int_{\{y^x : \|y^x - C(q)x\|^2\leq \lambda_{1}^w, y^x\in\mathbb{R}^{l}, x\in K_{i,q}\}} 1\,dy^x$ and $\beta_{2,q}^y = \phi_w^*\|C(q)\|$.  In other words, $\beta_{1,i,q}^y$ is the Lebesgue measure of region $K_{i,q}$ mapped to the observation space $\mathbb{R}^{l}$ via the matrix $C(q)$, and increased by $\sqrt{\lambda_{1}^w}$ in all directions, with $\lambda_{1}^w = \lambda_{\max}(\mathcal{W})$ the largest eigenvalue of $\mathcal{W}$.
\end{lem}

A similar result holds for the integral of $\tau_x$ over $\mathbb{R}^{m}$.
\begin{lem}\label{lem:finiteIntX}
For any $x, \overline{x}\in K_{i,q}$, for all $i=1,\dotsc,m_q$, $q\in\mathcal{Q}$, and any $u\in\mathcal{U}$, $q'\in\mathcal{Q}$, the following holds:
\begin{equation*}
\int_{\mathbb{R}^m} |\tau_x(x'|q',x,u) - \tau_x(x'|q',\overline{x},u)|\,dx 
\leq \left[\beta_{1,i,q}^xh_x^{(2)} + \beta_{2,q}^x\right] \delta_{i,q}^x,
\end{equation*}
$\beta_{1,i,q}^x = \int_{\{x' : \|x' - A(q)x-g(q,u)\|^2\leq \lambda_1^v, x'\in\mathbb{R}^{m}, x\in K_{i,q}\}} 1\,dx$ and $\beta_{2,q}^x = \phi_v^*\|A(q)\|$, with $\lambda_1^v$ the largest eigenvalue of $\mathcal{V}$.  
\end{lem}

In order to show convergence of  \eqref{eq:discreteViab} to \eqref{eq:optViab}, we require some additional definitions. First, similarly to $\hat{\sigma}_n$, we define piecewise constant function $\hat{\alpha}_n$ as $\hat{\alpha}_n(s) = \alpha_{n,\delta}(\xi(s))$, so that
\begin{equation*}
	\hat{\alpha}_n(s) = \int_{\sS}\int_{\sY}\hat{\alpha}_{n+1}^{*(y)}(s')\gamma(y|\xi(s'))\times\tau(s'|\xi(s),u){\bf 1}_{K_{\delta}}(\xi(s))\,dy\,ds'.
\end{equation*} 

We also define $\tilde{\alpha}_n(s)$ in the same manner as $\hat{\alpha}_n(s)$, except that it is directly related to $\alpha_n(s)$, i.e. uses the same optimal control input $u$, and the same combination of $\alpha_{n+1}$-functions (determined by $*(y)$).  In other words, $\tilde{\alpha}_{n}(s)$ is identical to $\alpha_n(s)$ in terms of the optimal policy tree from time $n$ to $N$, but the values are calculated using $\gamma(y|\xi(s'),u)$ and $\tau(s'|\xi(s),u)$,
\begin{equation}\label{eq:alphTilde1}
	\tilde{\alpha}_n^i(s) = \int_{\sS}\int_{\sY}\tilde{\alpha}_{n+1}^{i(y)}(s')\gamma(y|\xi(s'))\tau(s'|\xi(s),u^{i}){\bf 1}_{K_{\delta}}(\xi(s))\,dy,
\end{equation}
for a specific $\alpha$-function $i$ associated with $\alpha_n^i$. The superscript $i$ for $u^i$ and $i(y)$ indicates that the same choice of $u$ and combination of $\alpha_{n+1}^j(s)$ are used for both $\alpha_n^i(s)$ and $\tilde{\alpha}_n^i(s)$.  A bound on the difference between $\alpha_n^i(s)$ and $\tilde{\alpha}_n^i(s)$ is given in the following lemma.

\begin{lem}\label{lem:alphaTilde}
	For any $n\in[0,N]$, and any function $\alpha_n^i(s)\in\Gamma_n$, the associated function $\tilde{\alpha}_{n}^i(s)$ defined in \eqref{eq:alphTilde1} satisfies
	\begin{equation*}
		|\alpha_n^i(s) - \tilde{\alpha}_n^i(s)| \leq (N-n)N_q\left[\beta_1^yh_y^{(2)} + \beta_1^xh_x^{(2)} + \beta_2^y + \beta_2^x + h_q\right]\delta^x
	\end{equation*}
	for all $s\in\sS$.  The constants $\beta_1^y$ and $\beta_1^x$ are equal to $\max_{i=1,\dotsc,m_q, q\in\mathcal{Q}}\beta_{1,i,q}^y$ and $\max_{i=1,\dotsc,m_q, q\in\mathcal{Q}}\beta_{1,i,q}^x$ from Lemmas \ref{lem:finiteIntY} and \ref{lem:finiteIntX}, respectively.
\end{lem}

We now can show  convergence of the approximate safety probability over the discretized state space to the true safety probability.

\begin{thm}\label{thm:ValueFuncErrorDisc}
	For any time $n\in[0,N]$, and any $\sigma_n\in \Sigma_n$, $\sigma_{n,\delta}\in\Sigma_{n,\delta}$, the error between the value function \eqref{optDP}  and the value function \eqref{eq:valDiscrete} based on the finite  state approximation  is bounded above by
	\begin{equation*}
		\left|V_n^*(\sigma_n) - V_{n,\delta}^*(\sigma_{n,\delta})\right| \leq \eta_n^{\alpha}\delta^x
	\end{equation*}
	with $\eta_n^{\alpha} =N_q\lambda \eta_n^{\sigma} + (N-n)N_q(\beta_1^yh_y^{(2)} + \beta_1^xh_x^{(2)} + \beta_2^y + \beta_2^x)$.  
	
	Specifically, the safety probability for PODTSHS $\mathcal{H}$ over time horizon $N$ satisfies
	\begin{equation*}
		\left|p_{\mathrm{safe}}^N(\rho; K) - p_{\mathrm{safe,\delta}}^N(\rho_{\delta}; K_{\delta})\right| 
		\leq \left[N_qN(\beta_1^yh_y^{(2)} + \beta_1^xh_x^{(2)} + \beta_2^y + \beta_2^x)\right]\delta^x.
	\end{equation*}
\end{thm}

Theorem \ref{thm:ValueFuncErrorDisc} shows that the finite state approximation $\hat{\mathcal{H}}$ provides a means to approximately compute \eqref{eq:optViab} through the safety probability for $\hat{\mathcal{H}}$, \eqref{eq:discreteViab}.  As $\delta^x\rightarrow 0$, the finite state safety probability \eqref{eq:discreteViab} converges to the true value \eqref{eq:optViab}, and the policy $\pi_{\delta}^*$ converges to  $\pi^*$.

\begin{remark}
	The linearity and additive Gaussian noise assumptions for the continuous state dynamics and observations \eqref{stateUpdate} and \eqref{eq:yUpdate} are required only in the proofs of Lemmas \ref{lem:finiteIntY} and \ref{lem:finiteIntX}, and also in the sense that they guarantee the Lipschitz properties \eqref{eq:lipschitz} used in all proofs.  Therefore, nonlinear dynamics and non-Gaussian noise are permissible, so long as \eqref{eq:lipschitz} is satisfied, and an equivalent bound on the integrals in Lemmas \ref{lem:finiteIntY} and \ref{lem:finiteIntX} can be derived.  
\end{remark}

\subsection{Gaussian Mixture Approximation}\label{sec:abs_continuous}

We now consider a different approximation by representing the information state and $\alpha$-functions as Gaussian mixtures, such that each are characterized by a finite set of weights, means, and covariances, dependent on the mode $q$.  


Difficulty arises from the incorporation of the indicator function ${\bf 1}_K$ in \eqref{eq:alpha_yus} and \eqref{Phi}.  Integration over the compact set $K$ rather than all of $\mathcal{S}$ violates the preservation of the Gaussian form of $\sigma$ under operator $\Phi_{y,u}$, and similarly for the $\alpha$-functions.  To preserve the Gaussian mixture structure, we therefore propose a radial basis function (RBF) approximation \cite{Park91} to the indicator function, using Gaussians as the basis function.  For each $K_q$, we set
\begin{equation}\label{RBF_indApprox}
{\bf 1}_{K_q}(x) \approx \sum_{i=1}^{I_q} w_i(q) \phi(x; \mu_i(q), \sP_i(q)).
\end{equation}
 Typically, both the centers $\mu_i(q)$ and the number of components $I_q$ are fixed beforehand.  The weights and covariances can then be chosen to optimize the approximation. For simplicity we will denote $\phi(x; \mu_i(q), \sP_i(q))$ by $\phi_i(x)$.  This approximation is valid since the RBFs are dense in $L^p$ for $1\leq p \leq \infty$ \cite{Park91}, i.e. given any function $f$ in $L^p$, a weighted combination of RBFs can approximate $f$ to arbitrary accuracy given enough components, and ${\bf 1}_K$ is in $L^1$.

However, the discontinuity in ${\bf 1}_{K_q}$ produces the Gibbs phenomenon at the boundary of $K_q$ in the RBF approximation.  Although these oscillations will always exist for a finite number of components, they could  possibly be mitigated \cite{Fornberg11}.  The oscillations can be constrained to a smaller region of $K$ (shorter wavelength) with the addition of more components, indicating that the integrated error may be reduced even if the pointwise error is not.  Because we are interested only in integrating over $K$, this works to our advantage.


We define a new operator $\Phi^g$ and a new $\alpha$-function $\alpha_{y,u,\sigma}^g$ by inserting the RBF approximation \eqref{RBF_indApprox} into \eqref{Phi} and \eqref{eq:alpha_yus}, respectively.  We do not include the scaling factor $\frac{1}{p(y|\sigma,u)}$ in the operator $\Phi^g$, which is difficult to incorporate into a Gaussian mixture.  If the information state only represented the probability distribution over the current state of the system, we could simply ensure the weights all sum to one, but it includes the probability that past states were in $K$, and hence is not guaranteed to integrate to one.  Once we implement PBVI in Section IV, we are only concerned with finding the $\alpha$-function that maximizes the inner product $\langle \alpha_{n,g}, \sigma_{n,g}\rangle$, which is unaffected by a constant scaling factor, and hence we lose nothing by disregarding it.  We also presume continuous observations, as in Section \ref{sec:abs_finite}.   
\begin{align}
\left(\Phi_{y,u}^g\sigma_g\right)(s') &= \gamma(y|s')\sum_{q\in\mathcal{Q}}\int_{\mathbb{R}^m}\left[\sum_{i=1}^{I_q} w_i(q) \phi_i(x)\right]\tau(s'|s,u)\sigma_g(s)\,dx \label{eq:RBFPhi}\\
\alpha^g_{y,u,\sigma}(s) &= \int_{\mathcal{S}} \alpha_{n+1,g}^{*(y)}(s')\gamma(y|s')\tau(s'|s,u)\,ds'\left[\sum_{i=1}^{I_{q}} w_i(q) \phi_i(x) \right]\label{eq:RBFAlpha}
\end{align}

  We provide two lemmas stating that the operator $\Phi_{y,u}^g$ and  the $\alpha$-function update $\alpha_{y,u,\sigma}^g$ preserve the Gaussian mixture representation of $\sigma_{n,g}$ and $\alpha_{n,g}$ for all $n$.  These lemmas rely on the fact that $\tau_x$ and $\gamma_x$ are Gaussian, and that a product of Gaussian distributions is again a Gaussian.  We need some additional assumptions, however, to simplify the derivations and  make the following lemmas valid.

  \begin{assum}\label{ass:Gauss1}
    	The transition kernel $T_q$ does not depend on $x$, i.e. the mode update is dependent only on the previous mode: $T_q(q'|q,u)$.
  \end{assum}
  \begin{assum}\label{ass:Gauss3}
    	The matrices $A(q)$ and $C(q)$ in  \eqref{stateUpdate} and \eqref{eq:yUpdate} are invertible for all $q\in\sQ$.
 \end{assum}
 Without the first assumption, we would need to represent $T_q(q'|s,u)$ by a Gaussian mixture, which cannot be exact since we are approximating a function over discrete states by a continuous function \cite{Brunskill2010}.  The second assumption allows us to express $\tau_x(x'|q',x,u)$ and $\gamma_x(y|s)$ as Gaussian densities over $x$.  While these assumptions are quite restrictive, they can be relaxed at a cost of introducing additional error, as discussed later in Remark \ref{rem:Gauss}.

\begin{lem}\label{lem:PhiGauss}
The operator $\Phi_{y,u}^g$  is closed under Gaussian mixtures, i.e. for $\sigma_g$ a Gaussian mixture with $L$ components, $\Phi_{y,u}^g\sigma_g$ is also a Gaussian mixture with $N_qI_qL$ components for any $u\in\mathcal{U}$, $y\in\mathcal{Y}$.
\end{lem}

\begin{lem}\label{lem:alphaGauss}
The expression for $\alpha_{y,u,\sigma}^g$ is closed under Gaussian mixtures, i.e. if $\alpha_{n+1,g}^{*(y)} $ is a  Gaussian mixture with $M$ components, $\alpha^g_{y,u,\sigma}$ is also a Gaussian mixture with $N_qI_qM$ components, for any $u\in\mathcal{U}$, $y\in\mathcal{Y}$, $\sigma\in\Sigma$.
\end{lem}

 Lemma \ref{lem:PhiGauss} implies that we can approximate $\sigma$ through a Gaussian mixture and use the equivalent update operator $\Phi_{y,u}^g$, hence the Gaussian mixture approximation of $\sigma$ is $\sigma_{n,g}(x,q) = \sum_{l=1}^L w^{\sigma}_{l,n}(q)\phi(x; \mu^{\sigma}_{l,n}(q), \sP^{\sigma}_{l,n}(q))$.  Similarly, the Gaussian mixture of any $\alpha$-function is written as $\alpha_{n,g}(x,q) =\sum_{m=1}^Mw_{m,n}^{\alpha}\phi(x; \mu_{m,n}^{\alpha}(q), \sP_{m,n}^{\alpha}(q))$ for all $n$.
\begin{equation}
\begin{aligned}
\sigma_{0,g}(x,q) &= \sigma_0(x,q) = R_q(q)\phi(x; \mu_0; \sP_0) \\
\sigma_{n,g}(x,q) &= \sum_{l=1}^L w^{\sigma}_{l,n}(q)\phi(x; \mu^{\sigma}_{l,n}(q), \sP^{\sigma}_{l,n}(q)) 
\end{aligned}.
\end{equation}
Similarly, the Gaussian mixture approximation of any $\alpha$-function is written:
\begin{equation}
\begin{aligned}
\alpha_{N,g}(x,q) &=\sum_{i=1}^{I_{q}} w_i(q) \phi_i(x) \\
\alpha_{n,g}(x,q) &=\sum_{m=1}^Mw_{m,n}^{\alpha}\phi(x; \mu_{m,n}^{\alpha}(q), \sP_{m,n}^{\alpha}(q))
\end{aligned}.
\end{equation}
The weights, means, and covariances are defined recursively.  Their exact representations can be found in the Appendix.

Note that although the Gaussian mixture representation of $\alpha_{y,u,\sigma}^g$ has a finite number of components given that the representation of $\alpha_{n+1,g}$ is finite, the actual $\alpha$-function, $\alpha_{n,g}$, is expressed as the integral of $\alpha_{y,u,\sigma}^g$ over $\mathcal{Y}$.  Therefore, without the assumption that $\mathcal{Y}$ is finite, $\alpha_{n,g}$ must have an infinite number of components (by breaking the integral over $\mathcal{Y}$ into a summation over regions of size $\Delta \subset \mathcal{Y}$ and taking the limit as $\Delta\rightarrow 0$).  We take some liberty in overlooking this discrepancy, because it does not affect the proofs in this section.  We impose a finite set $\mathcal{Y}$ in Section \ref{sec:pbvi}, which makes the Gaussian mixture representation of the $\alpha$-functions indeed valid, and discuss additional error implications.

The subscript $g$ denotes that we are computing an estimate based on the Gaussian mixture approximation \eqref{RBF_indApprox}.  The value function $V_{n,g}^*(\sigma_{n,g})$ is described through the recursion
\begin{equation}\label{eq:valGauss}
\begin{aligned}
V_{N,g}^*(\sigma_{N,g}) &= \sum_{\mathcal{Q}}\int_{\mathbb{R}^{m}}\sum_{i=1}^{I_q}w_i(q)\phi_i(x)\sigma_{N,g}(x,q)\,dx \\
V_{n,g}^*(\sigma_{n,g}) &= \max_{u\in\mathcal{U}}\int_{\mathcal{Y}}V_{n+1,g}^*(\Phi_{y,u}^g\sigma_{n,g})\mathbb{P}(dy|\sigma_{n,g},u)
\end{aligned}.
\end{equation}

The safety problem  for the Gaussian mixture approximation is defined as 
\begin{equation}\label{eq:gaussViab}
 p_{\mathrm{safe},g}^N(\rho; K) 
= V_{0,g}^*(\rho).
\end{equation}

Since $\tau_x$, $\gamma_x$, and $\rho_x$ are Gaussian, and based on Assumptions \ref{ass:Gauss1} - \ref{ass:Gauss3} made above, the Gaussian mixture representations of $\alpha$ and $\sigma$ are exact, aside from the approximation of ${\bf 1}_K$ using RBFs. To quantify the error incurred from calculating $V_{0,g}^*$ as opposed to $V_0^*$ (from integration of \eqref{RBF_indApprox} over $\mathcal{S}$ rather than over $K$), we define the error
\begin{equation}\label{eq:L2IndError}
\delta^{I}=  \max_{q\in\sQ}\left\|{\bf 1}_{K_q} - \sum_{i=1}^{I_q}w_i(q)\phi_i(x)\right\|_1.
\end{equation}



We first analyze the error between $\sigma_n$ and $\sigma_{n,g}$, which is stated in terms of the 1-norm rather than the $\infty$-norm, to be consistent with considering the integrated error in \eqref{eq:L2IndError}.

\begin{thm}\label{thm:sigContError}
The Gaussian mixture approximation $\sigma_{n,g}$ of $\sigma_n$ satisfies 
\begin{equation*}
\left\|\sigma_n - \sigma_{n,g}\right\|_{1} \leq \gamma_n^{\sigma} N_q\delta^I
\end{equation*}
for any $n\in[0,N]$, $y\in\mathcal{Y}$, and $u\in\mathcal{U}$, with $\gamma_n^{\sigma} = \displaystyle\sum_{j=0}^{n-1}(\phi_w^*)^{j+1}\phi_{\sigma,j}^*$ and $\phi_{\sigma, j}^*=\displaystyle\max_{l\in1,\dotsc,L; q\in\mathcal{Q}}(2\pi)^{-\frac{m}{2}}|\Sigma_{l,j}^{\sigma}(q)|^{-\frac{1}{2}}$.
\end{thm}


To show convergence of \eqref{eq:gaussViab} to \eqref{eq:optViab},  we define the function $\tilde{\alpha}^i_{n,g}(s)$ which utilizes the same policy tree as $\alpha^i_n(s)$ for a specific $\alpha_{n}^i(s)\in\Gamma_n$.  This is equivalent to \eqref{eq:alphTilde1}, except that $\tilde{\alpha}^i_{n,g}(s)$ is defined by replacing ${\bf 1}_K(s)$ with \eqref{RBF_indApprox}.
\begin{equation}\label{eq:alphGtil}
\tilde{\alpha}_{n,g}^i(s) = \int_{\sS}\int_{\sY}\tilde{\alpha}_{n+1,g}^{i(y)}(s')\gamma(y|s')\tau(s'|s,u^i)\,dy\,ds'
\left[\sum_{i=1}^{I_q}w_i(q)\phi_i(x)\right]
\end{equation}
with $u^i$ the optimal control input associated with $\alpha_n^i(s)$ and $i(y)$ indicating that $\alpha^{i(y)}_{n+1,g}(s)$ is chosen according to the indices selected by $*(y)$ for $\alpha_{n}(s)$.  

Note that the $\alpha$-functions are no longer guaranteed to be equal to zero outside of $K$, and also are not guaranteed to be bounded above by one.  So long as \eqref{RBF_indApprox} is of bounded height, however, the $\alpha$-functions also remain bounded, and we write $\max_{\alpha_{n,g}\in\Gamma_{n,g}}\|\alpha_{n,g}\|_{\infty} = \overline{\alpha}_n$.  We could also adjust the weights in \eqref{RBF_indApprox} so that the $\alpha$-functions cannot exceed one, although this may increase the error $\delta^I$.

The following lemma describes the relation between $\alpha_n(s)$ and $\tilde{\alpha}_n(s)$.
\begin{lem}\label{lem:alphGtil}
	For any $n\in[0,N]$, and any $\alpha_n^i(s)\in\Gamma_n$, the associated function $\tilde{\alpha}_{n,g}^i(s)$ defined in \eqref{eq:alphGtil} satisfies
	\begin{equation*}
	\left\|\alpha_n^i - \tilde{\alpha}^i_{n,g}\right\|_{1} \leq \left(\sum_{k=n}^N(\lambda\phi_v^*)^{N-k}\overline{\alpha}_{N-k+n+1}\right)N_q\delta^I
	\end{equation*}
	where $\overline{\alpha}_{N+1} = 1$.
\end{lem}

\begin{thm}\label{thm:ValueFuncErrorGauss}
	For any time $n\in[0,N]$, and any $\sigma_n\in \Sigma_n$, $\sigma_{n,g}\in\Sigma_{n,g}$, the error between the value function \eqref{optDP} given $\sigma_n$ and the value function \eqref{eq:valGauss} given $\sigma_{n,g}$ using the Gaussian mixture approximation  is bounded above by
	\begin{equation*}
	\left|V_n^*(\sigma_n) - V_{n,g}^*(\sigma_{n,g})\right| \leq \gamma^{\alpha}_nN_q\delta^I
	\end{equation*}
	with $\gamma^{\alpha}_n = \sum_{k=n}^N(\lambda\phi_v^*)^{N-k}\overline{\alpha}_{N-k+n+1}\phi_{\sigma,n}^*\delta^I +  \overline{\alpha}_n\gamma_n^{\sigma}$.
	
	Specifically, the safety probability for PODTSHS $\mathcal{H}$ over time horizon $N$ satisfies
	\begin{equation*}
	\left|p_{\mathrm{safe}}^N(\rho; K) - p_{\mathrm{safe},g}^N(\rho; K)\right| \leq \gamma_0^{\sigma} \delta^I.
	\end{equation*}
\end{thm}

Theorem \ref{thm:ValueFuncErrorGauss} shows that the convergence of the Gaussian mixture approximation of both $\sigma$ and the value function depends on the integrated error between the indicator function over $K$ and the RBF approximation \eqref{RBF_indApprox}, rather than the pointwise error.  Although the pointwise error may not converge to zero for a finite number of components in the RBF, the integral of the error can be small, as we will show in Section \ref{sec:example}.

\begin{remark}\label{rem:Gauss}
	The linearity and additive Gaussian noise assumptions on the dynamics \eqref{stateUpdate} and \eqref{eq:yUpdate}, as well as Assumptions \ref{ass:Gauss1} - \ref{ass:Gauss3},  are used in the Gaussian mixture approximation to ensure that the only error in the value function and information state approximations comes from the approximation of the indicator function ${\bf 1}_K(s)$ by a Gaussian mixture.  Dropping these assumptions requires that we approximate $\tau_x$, $T_q$, and $\gamma_x$ by Gaussian mixtures, which is possible but introduces additional error that we have chosen not to consider. 
\end{remark}

\section{Approximate Numerical Solution with Lower Bound}\label{sec:pbvi}

A numerical solution of Problem \ref{probState} via either a discrete or Gaussian mixture approximation additionally requires sets $\Gamma_n$ and $\Sigma_n$ to be finite, whereas we have sets of infinite size because of the uncountable nature of $\mathcal{Y}$.  However, a lower bound on the safety probabilities \eqref{eq:discreteViab} and \eqref{eq:gaussViab} can still be obtained, by characterizing the error that results from using $\tilde{\Gamma}_n\subset\Gamma_n$ and $\tilde{\Sigma}_n\subset\Sigma_n$, finite collections of $\alpha$-functions and information states, respectively.

We again exploit  point-based approximation methods described in Section \ref{sec:back_pomdp}. We examine the generation of subsets of the information states and $\alpha$-functions, and prove that each guarantees a lower bound to the safety probability of whichever approximation of Section \ref{sec:abstraction} we choose.    In contrast to most point-based solvers, we do not assume a finite set of observations, and hence discretize the observations merely for the computation of the $\alpha$-functions. Combining belief space sampling with discretized observations assures a lower bound to the safety probability.


\subsection{Sampling from the information space}

We  characterize the error from using a sampled subset of $\Sigma_n$ for performing backup operations (as in \eqref{backup}).  Presume that a finite set of information states $\tilde{\Sigma}_n$ has been generated according to one of the many methods available \cite{shani13}.  We  generate a finite set $\tilde{\Gamma}_n$ of $\alpha$-functions, one for each $\sigma_n \in \tilde{\Sigma}_n$.  The convexity of the value functions guarantees that the subset $\tilde{\Gamma}_n$  provides a lower bound on $V_n^*$.  Further, we can show that the error between the approximate value function $\tilde{V}_n^*$ (represented by $\tilde{\Gamma}_n$) and the true value function $V_n^*$ (represented by the complete set of $\alpha$-functions $\Gamma_n$) depends on how densely we sample $\Sigma_n$.

We  define an intermediate value function $\hat{V}_n^*(\sigma_n) = \sup_{\hat{\alpha}_n\in\tilde{\Gamma}_n}\langle \hat{\alpha}_n, \sigma_n \rangle$ that generates  $\tilde{\Gamma}_n$ recursively from $\Gamma_{n+1}$, i.e. that introduces one point-based backup from the full set $\Gamma_{n+1}$.  Then $\hat{\alpha}_n$ is written as a function of $\alpha_{n+1}^*$ rather than $\tilde{\alpha}_{n+1}^*$.

The value function $\tilde{V}_n^*$ is formally defined as $\tilde{V}_n^*(\sigma_n) = \sup_{\tilde{\alpha}_n\in\tilde{\Gamma}_n}\langle \tilde{\alpha}_n, \sigma_n \rangle$ 
\begin{equation}\label{eq:valFuncTilde}
\begin{aligned}
\tilde{\alpha}_n(s) &= \int_{\mathcal{Y}}\int_{\mathcal{S}}\tilde{\alpha}_{n+1}^{*(y)}(s')\gamma(y|s')\tau(s'|s,u){\bf 1}_K(s)\,dy\,ds'\\
\tilde{\alpha}_{n+1}^{*(y)}(s') &= \arg\left\{\sup_{\tilde{\alpha}_{n+1}\in\tilde{\Gamma}_{n+1}}\int_{\mathcal{S}}\tilde{\alpha}_{n+1}(s')\gamma(y|s')\int_K\tau(s'|s,u)\sigma_n(s)\,ds'\,ds\right\}
\end{aligned}
\end{equation}
so that $\tilde{V}_n^*$ is characterized by the finite set $\tilde{\Gamma}_n$ at each time step.  We also define an intermediate value function $\hat{V}_n^* = \sup_{\hat{\alpha}_n\in\tilde{\Gamma}_n}\langle \hat{\alpha}_n, \sigma \rangle$ that generates  $\tilde{\Gamma}_n$ recursively from $\Gamma_{n+1}$, i.e. that introduces one point-based backup from the full set $\Gamma_{n+1}$.  Then $\hat{\alpha}_n$ is written as a function of $\alpha_{n+1}^*$ rather than $\tilde{\alpha}_{n+1}^*$.
\begin{equation}
\label{eq:valFuncHat}
\alpha_{n+1}^{*(y)}(s') = \arg\left\{\sup_{\alpha_{n+1}\in\Gamma_{n+1}}\int_{\mathcal{S}}\alpha_{n+1}(s')\gamma(y|s')\int_K\tau(s'|s,u)\sigma_n(s)\,ds'\,ds\right\}.
\end{equation}
 
 We let $\delta^{\sigma}$ denote the maximum Hausdorff distance over $n$ between points in $\tilde{\Sigma}_n$ and points in $\Sigma_n$ with respect to the metric $\|\cdot\|_1$.
\begin{equation}\label{delta}
\delta^{\sigma} = \max_n\left\{\max\{\sup_{\tilde{\sigma}_n\in\tilde{\Sigma}_n}\inf_{\sigma_n\in\Sigma_n}\|\tilde{\sigma}_n - \sigma_n\|_1, \sup_{\sigma_n\in\Sigma_n}\inf_{\tilde{\sigma}_n\in\tilde{\Sigma}_n}\|\tilde{\sigma}_n - \sigma_n\|_1\}\right\}
\end{equation}
In the following, we do not distinguish between the vector and Gaussian mixture representations of $\sigma$ and $\alpha$, because the results apply to both cases. 

\begin{lem}\label{lemma3}
For any $n\in[0,N]$ and $\sigma_n\in\Sigma_n$, the error introduced in one iteration of point-based value iteration is at most $\delta^{\sigma}$.
\begin{equation*}
 |\hat{V}_n^*(\sigma_n) - V_n^*(\sigma_n)| \leq \delta^{\sigma}
\end{equation*}
\end{lem}

We now use Lemma \ref{lemma3} to derive a bound between the true value function and the point-based approximation at any time $n$.  

\begin{thm}\label{thm1}
For a set of information states $\Sigma_n$, sampled set $\tilde{\Sigma}_n$, and any time $n\in[0,N]$ and  $\sigma_n\in\Sigma_n$, the error from using point-based value iteration versus full value iteration is bounded above by
\begin{equation*}
 |\tilde{V}_n^{*}(\sigma_n) - V_n^*(\sigma_n)|\leq (N-n) \delta^{\sigma}.
\end{equation*}
\end{thm}

Thus the error between the point-based approximation and the actual value function is directly proportional to how densely $\tilde{\Sigma}_n$ is sampled, and converges to zero as $\tilde{\Sigma}_n$ approaches $\Sigma_n$.  The proofs of Lemma \ref{lemma3} and Theorem \ref{thm1} are a straightforward extension of those appearing in \cite{pineau06}, and are omitted.

\subsection{Calculating the $\alpha$-functions}

Over the uncountably infinite space $\mathcal{Y}$, we cannot calculate $\alpha_{y,u,\sigma}$ for all $y\in\mathcal{Y}$, despite a finite set of $u$ and $\sigma$.  We therefore compute a subset of $\alpha_{y,u,\sigma}$ for the finite set $y^i$, to approximate $\alpha_n$ as
$\alpha_n(s) \approx \sum_{y^i}\alpha_{y^i,u,\sigma}(s)$.  We discretize $\mathcal{Y}$ in a similar fashion to the discretization of $\mathcal{S}$ in Section \ref{sec:abs_finite}.  

However, since $\mathbb{R}^{l}$ is not compact, we consider an expanded set  $\overline{K} =\bigcup_{y^q\in\mathcal{Q}} \overline{K}_{y^q}\supset K$ defined so that the probability of observing a value $y$ for $s\in K$ that is outside of $\overline{K}$ is approximately zero, i.e. $\gamma(\mathcal{Y}\backslash \overline{K}|s\in K) < \epsilon$, $\epsilon\ll 1$.   The sets $\overline{K}_{y^q}$ are divided into disjoint subsets $\overline{K}_{i,y^q}$, $\bigcup_{i = 1,\dotsc l_q}\overline{K}_{i,y^q} = \overline{K}_{y^q}$.  We also define $\psi_y= \overline{K}^c = \mathcal{Y}\backslash \overline{K}$, such that $\bigcup_{i=1,\dotsc,l_q}\overline{K}_{i,y^q} \times \{\psi_y\}= \mathbb{R}^n$.

The partition of $\overline{K}$ is denoted $\mathcal{G}^y = \bigcup_{i,y^q} \mathcal{G}_{i,y^q}^y$ with $\mathcal{G}_{i,y^q}^y = \{\overline{K}_{i,y^q}\times y^q  : i = 1,\dotsc l_q, y^q\in\mathcal{Q}\}$.  The diameter of partition $\overline{K}_{i,y^q}$ is $\delta_{i,y^q}^y=\sup\{\|y-\overline{y}\| : y, \overline{y}\in \overline{K}_{i,y^q}\}$, with maximum diameter $\delta^y = \max_{i,y^q} \delta_{i,y^q}^y$.  Each partition $\mathcal{G}_{i,y^q}^y$ has a representative element $(y^{x,i,y^q}, y^q)$ and a set $\mathcal{Y}_{\delta} = \{( y^{x,i,y^q},y^q) :  i=1,\dotsc,l_q, y^q\in\mathcal{Q}\}$.  The function $\theta: \mathcal{Y}\rightarrow \mathcal{Y}_{\delta}$ maps observation $y\in\mathcal{Y}$ to its representative value $(y^{x,i,y^q}, y^q)$; the set-valued function $\Theta: \mathcal{Y}_{\delta}\rightarrow \overline{K}$ maps the point $( y^{x,i,y^q}, y^q)$ to the corresponding set $\mathcal{G}_{i,y^q}$.

The finite observation space is $W_{\delta} = \mathcal{Y}_{\delta}\bigcup \{\psi_y\}$.  For the finite state approximation, the transition function  $\gamma_{\delta}: W_{\delta}\times K_{\delta} \rightarrow [0,1]$ is defined as
\begin{equation}\label{discreteYker}
\gamma_{\delta}(w|z) = \begin{cases}
Y(\Theta(w)|z), &\text{if $w \in \mathcal{Y}_{\delta}$} \\
1 - \sum_{\overline{w}\in \mathcal{Y}_{\delta}} Y(\Theta(\overline{w})|z), &\text{if $w = \psi_y$} \\
\end{cases}.
\end{equation}

For the Gaussian mixture approximation, we define the transition function $\gamma_g$ in the same fashion as \eqref{discreteYker}, but with 
\begin{equation}\label{gaussYker}
Y(\Theta(w)|z)\approx Y_q(y^q|q)\sum_{j=1}^{M_y}c_j\phi_j^y(y^{x,i,y^q}_j; C(q)x, \mathcal{W})
\end{equation}
so that the $\alpha$-functions will also be Gaussian mixtures at each time step.  Note that $w = (y^{x,i,y^q},y^q)$, $y^{x,i,y^q}_j$ is a set of mesh points inside $\mathcal{G}_{i,y^q}$ associated with $w$, and $c_j$ is a weight proportional to the mesh spacing (determined, e.g., by the trapezoidal rule for numerical integration).

\subsubsection{Discretized observations for finite state approximation}

We use $\tilde{\Gamma}_{n,\delta}$ and $\tilde{V}_{n,\delta}^*$ to denote the approximation using a finite subset of $\Gamma_{n,\delta}$, with the important distinction that the subset is now generated by a finite collection of observations (as opposed to $\tilde{\Sigma}_n$, i.e. we assume here that the set $\Sigma_n$ is finite).

The value function  is then $\sup_{\tilde{\alpha}_{n,\delta}\in\tilde{\Gamma}_{n,\delta}}\sum_{z\in K_{\delta}} \tilde{\alpha}_{n,\delta}(z)\sigma_{n,\delta}(z)$, where $\tilde{\alpha}_{n,\delta}$ defined as in Section \ref{sec:abs_finite}, only with $\gamma(y|z)$ replaced by $\gamma_{\delta}(w|z)$.
\begin{equation*}\tilde{V}^*_{n,\delta}(\sigma_{n,\delta}) =  \sup_{\tilde{\alpha}_{n,\delta}\in\tilde{\Gamma}_{n,\delta}}\sum_{z\in K_{\delta}} \tilde{\alpha}_{n,\delta}(z)\sigma_{n,\delta}(z),\end{equation*}
 with
\begin{equation}
\label{eq:valDiscTilde}
\begin{aligned}
\tilde{\alpha}_{n,\delta}(z) &= \sum_{w\in W_{\delta}}\sum_{z'\in K_{\delta}}\tilde{\alpha}_{n+1,\delta}^{*(w)}(z')\gamma_{\delta}(w|z')\tau_{\delta}(z'|z,u)\\
\tilde{\alpha}_{n+1,\delta}^{*(w)}(z') &= \arg\left\{\sup_{\tilde{\alpha}_{n+1,\delta}\in\tilde{\Gamma}_{n+1,\delta}}\sum_{z'\in K_{\delta}}\tilde{\alpha}_{n+1,\delta}(z') \gamma_{\delta}(w|z')\tau_{\delta}(z'|z,u)\sigma_{n,\delta}(z)\right\}.
\end{aligned}
\end{equation}
We again define the intermediate value function $\hat{V}_{n,\delta}^*(\sigma_{n,\delta})$, with $\hat{\alpha}_{n,\delta}$ calculated using $\alpha_{n+1,\delta}^{*(w)}\in \Gamma_{n+1,\delta}$ (as opposed to $\tilde{\Gamma}_{n+1,\delta}$) to capture the error introduce in one backup iteration using discretized observations.

We can then bound the error introduced in one iteration of approximating the $\alpha$-vectors through discretized observations.
\begin{lem}\label{lem:discObsAlpha}
For any time $n\in[0,N]$ and $\sigma_{n,\delta}\in\Sigma_{n,\delta}$, the error between $V_{n,\delta}^*(\sigma_{n,\delta})$ and $\hat{V}_{n,\delta}^*(\sigma_{n,\delta})$ satisfies 
\begin{equation*}
0\leq V_{n,\delta}^*(\sigma_{n,\delta}) - \hat{V}_{n,\delta}^*(\sigma_{n,\delta}) \leq N_q\overline{\lambda}h_y^{(1)}\delta^y +\frac{\epsilon}{N}
\end{equation*}
given that the discretized observations $w$ are chosen so that 
\begin{equation*}
\gamma_{\delta}(w|z') > \gamma(w|z')|\Theta(w)|,
\end{equation*}
and with $\overline{\lambda}$ the largest Lebesgue measure of sets $\overline{K}_{y^q}$.  
\end{lem}

Lemma \ref{lem:discObsAlpha} requires defining the representative points $w = (y^{x,i,y^q},y^q)$ so that the integral of $\gamma$ over $\Theta(w)$ is greater than a piecewise constant approximation integrated over $\Theta(w)$, which can be satisfied by picking the points $y^{x,i,y^q}$ where the Gaussian density $\gamma_x$ is minimized within cell $\overline{K}_{i,y^q}$.  Without this requirement, finding $\alpha_{n+1,\delta}^{*(w)}$ at a finite set of points still guarantees a lower bound to the value function for any time $n$, and is intuitively more accurate as $\delta^y\rightarrow 0$.

Lemma \ref{lem:discObsAlpha} leads to the following theorem regarding the error between $V_{n,\delta}^*(\sigma_{n,\delta})$ (based on continuous observations) and $\tilde{V}^*_{n,\delta}(\sigma_{n,\delta})$ (based on discretized observations).  We again use the notation $\tilde{V}$ to indicate that $\tilde{V}$ is represented by the set $\tilde{\Gamma}$ of $\alpha$-functions calculated using the discretized observations.
\begin{thm}\label{thm:discObsVF}
Given discretized observation process $W_{\delta}$ with transition function \eqref{discreteYker}, for any time $n\in[0,N]$, the error between $V_{n,\delta}^*(\sigma_{n,\delta})$ calculated according to $\mathcal{Y}$ and  $\tilde{V}^*_{n,\delta}(\sigma_{n,\delta})$ calculated according to $W_{\delta}$ satisfies
\begin{equation*}
0\leq V_{n,\delta}^*(\sigma_{n,\delta})- \tilde{V}^*_{n,\delta}(\sigma_{n,\delta}) \leq (N-n)N_q\overline{\lambda}h_y^{(1)}\delta^y + \frac{(N-n)\epsilon}{N}
\end{equation*}
for any $\sigma_{n,\delta}\in\Sigma_{n,\delta}$, with $\overline{\lambda}$ the largest Lebesgue measure of sets $\overline{K}_{y^q}$.  

Specifically, the safety probability for  $\hat{\mathcal{H}}$ satisfies
\begin{equation*}
0\leq p_{\mathrm{safe},\delta}^N(\rho_{\delta}; K_{\delta}) - \tilde{V}_{0,\delta}^*(\rho_{\delta}) \leq NN_q\overline{\lambda}h_y^{(1)}\delta^y + \epsilon.
\end{equation*}
\end{thm}

\subsubsection{Discretized observations for Gaussian mixture approximation}

The results of discretizing the observations for the Gaussian mixture abstraction are nearly identical to those for the finite state abstraction.  The main difference arises in approximating the integral $Y(\Theta(w)|s')$ with a Gaussian sum. To ensure the approximate value function provides a lower bound to $V_{n,g}^*$, we must under-approximate the integral $Y(\Theta(w))$ for each $w$. We define $\tilde{V}_{n,g}^*$ similarly to $\tilde{V}_{n,\delta}^*$, using discretized observations and the RBF approximation to the indicator function,
\begin{equation}
\label{eq:valTildeGauss}
\begin{aligned}
\tilde{\alpha}_{n,g}(s) &= \sum_{w\in W_{\delta}}\int_{\mathcal{S}}\tilde{\alpha}_{n+1,g}^{*(w)}(s')\gamma_g(w|s')\tau(s'|s,u)\,ds'\sum_{i=1}^{I_q}w_i(q)\phi_i(x) \\
\tilde{\alpha}_{n+1,g}^{*(w)}(s') &= \arg\left\{\sup_{\tilde{\alpha}_{n+1,g}\in\tilde{\Gamma}_{n+1,g}}\sum_{q\in\mathcal{Q}}\int_{\mathbb{R}^m}\tilde{\alpha}_{n+1,g}^{*(w)}(s')\gamma_g(w|s')\tau(s'|s,u)\,ds'\left[\sum_{i=1}^{I_q}w_i(q)\phi_i(x)\right]\sigma_{n,g}(x,q)\,dx\right\}
\end{aligned}
\end{equation}
and $\hat{V}_{n,g}^*$ is the intermediate value function that finds the optimal $\alpha_{n+1,g}^{*(w)}\in \Gamma_{n+1,g}$, rather than in $\tilde{\Gamma}_{n+1,g}$.  We can bound the error between $V_{n,g}^*$ and $\hat{V}_{n,g}^*$, and between $V_{n,g}^*$ and $\tilde{V}_{n,g}^*$, equivalently to Lemma \ref{lem:discObsAlpha} and Theorem \ref{thm:discObsVF}, respectively.
\begin{lem}\label{lem:gaussObsAlpha}
	For any time $n\in[0,N]$ and $\sigma_{n,g}\in\Sigma_{n,g}$, the error between $V_{n,g}^*(\sigma_{n,g})$ and $\hat{V}_{n,g}^*(\sigma_{n,g})$ satisfies 
	\begin{equation*}
	0\leq V_{n,g}^*(\sigma_{n,g}) - \hat{V}_{n,g}^*(\sigma_{n,g}) \leq N_q\overline{\lambda}h_y^{(1)}\overline{\alpha}_{n+1}\delta^y +\frac{\epsilon}{N}
	\end{equation*}
	given that the observations $w$ are chosen so that 
	\begin{equation*}
 \gamma_{g}(w|s')
 > \gamma(w|s')|\Theta(w)|,
	\end{equation*}
	and with $\overline{\lambda}$ the largest Lebesgue measure of sets $\overline{K}_{y^q}$.  
\end{lem}
\begin{thm}\label{thm:gaussObsVF}
	Given discretized observation process $W_{\delta}$ with transition function \eqref{gaussYker}, for any time $n\in[0,N]$, the error between $V_{n,g}^*(\sigma_{n,g})$ calculated according to $\mathcal{Y}$ and  $\tilde{V}^*_{n,g}(\sigma_{n,g})$ calculated according to $W_{\delta}$ satisfies
	\begin{equation*}
	0\leq V_{n,g}^*(\sigma_{n,g})- \tilde{V}^*_{n,g}(\sigma_{n,g}) \leq N_q\overline{\lambda}h_y^{(1)}\left(\sum_{i=n+1}^N\overline{\alpha}_i\right)\delta^y+ \frac{(N-n)\epsilon}{N}
	\end{equation*}
	for any $\sigma_{n,g}\in\Sigma_{n,g}$, with $\overline{\lambda}$ the largest Lebesgue measure of sets $\overline{K}_{y^q}$.  
	
	Specifically, the safety probability for the Gaussian mixture approximation satisfies
	\begin{equation*}
	0\leq p_{\mathrm{safe},g}^N(\rho; K) - \tilde{V}_{0,g}^*(\rho) \leq N_q\overline{\lambda}h_y^{(1)}\left(\sum_{i=1}^N\overline{\alpha}_i\right)\delta^y + \epsilon.
	\end{equation*}
\end{thm}

The proofs of Lemma \ref{lem:gaussObsAlpha} and Theorem \ref{thm:gaussObsVF} follow directly from the proofs of Lemma \ref{lem:discObsAlpha} and Theorem \ref{thm:discObsVF}.

To summarize, given either the finite state or Gaussian mixture approximation, we can subsequently 1) sample $y$ from $\mathcal{Y}$ and $u$ from $\mathcal{U}$ to generate the progressive subsets $\tilde{\Sigma}_{n,\delta}$ or $\tilde{\Sigma}_{n,g}$, and 2) discretize $\mathcal{Y}$ and use the set $W_{\delta}$ to calculate $\tilde{\alpha}_{w,u,\sigma_{\delta}}^{\delta}$ or $\tilde{\alpha}_{w,u,\sigma_g}^g$, which are then used to generate $\tilde{\alpha}_{n,\delta}\in\tilde{\Gamma}_{n,\delta}$ and $\tilde{\alpha}_{n,g}\in\tilde{\Gamma}_{n,g}$.  Using sets $\tilde{\Sigma}_{n,\delta}$ and $\tilde{\Gamma}_{n,\delta}$ in place of $\Sigma_{n,\delta}$ and $\Gamma_{n,\delta}$  provides a lower bound to the safety probability $p_{\mathrm{safe},\delta}^N(\rho_{\delta}; K_{\delta})$ that converges as $\delta^{\sigma}$ and $\delta^y$ approach zero (and similarly for $\tilde{\Sigma}_{n,g}$ and $\tilde{\Gamma}_{n,g}$).

\begin{figure*}[!ht]
	\centering
	\subfloat[]{\label{fig:discreteXcompare}\includegraphics[width=.4\textwidth,keepaspectratio]{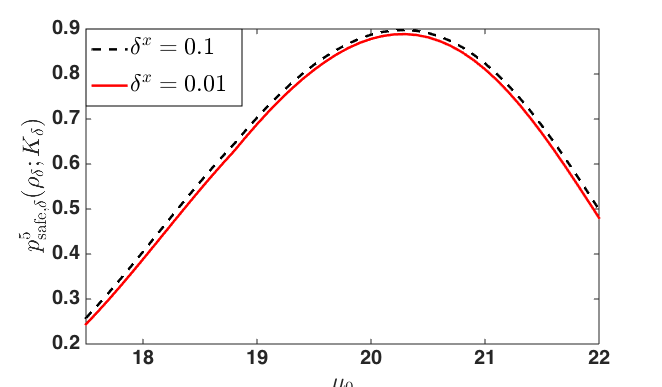}}\quad
	\subfloat[]{\label{fig:gaussIndcompare}\includegraphics[width=.4\textwidth]{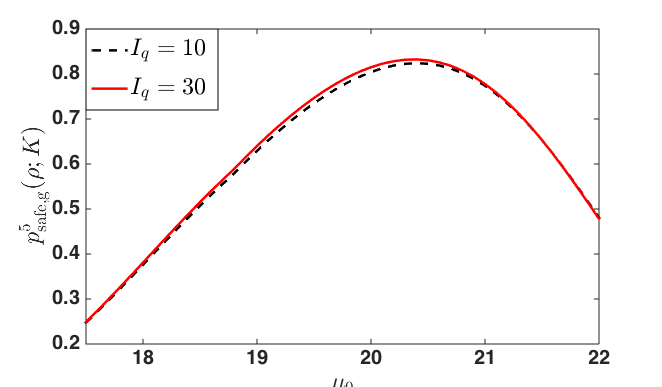}}
	\caption{\small{Comparison of safety probabilities over varying initial distribution $\rho = \phi(x; \mu_0, 1)$ and $q_0 = 0$, using the finite state approximation (a) and Gaussian mixture approximation (b).  In both (a) and (b) $\delta^y = 0.5$.  Fig. (a) compares probabilities for $\delta^x = 0.1$ (black dashed line) and $\delta^x = 0.01$ (red solid line). Fig. (b) compares probabilities for $I_q = 10$ (black dashed line) and $I_q = 30$ (red solid line).  The refinement of $\delta^x$ and increase in $I_q$ have a small impact on the safety probabilities.  
		}}
		\label{fig:viabCompare}
		
	\end{figure*}
	
	\begin{figure*}[!th]
		\centering
		\subfloat[]{\label{fig:discreteXcompareU}\includegraphics[width=.4\textwidth]{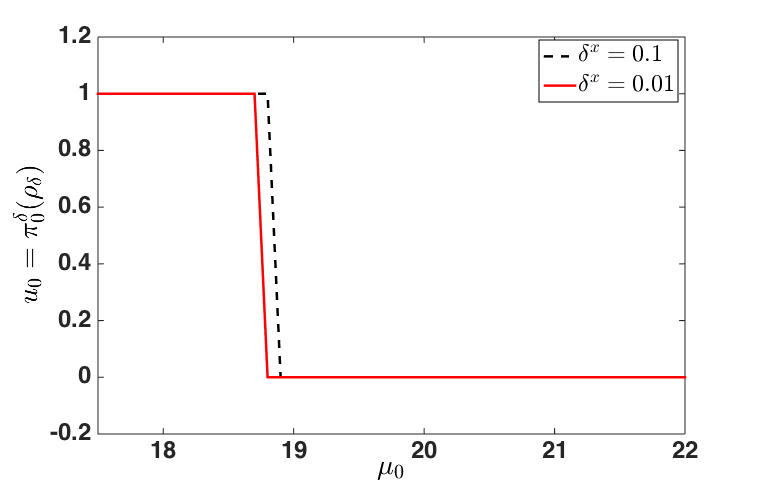}}\quad
		\subfloat[]{\label{fig:gaussIndcompareU}\includegraphics[width=.4\textwidth]{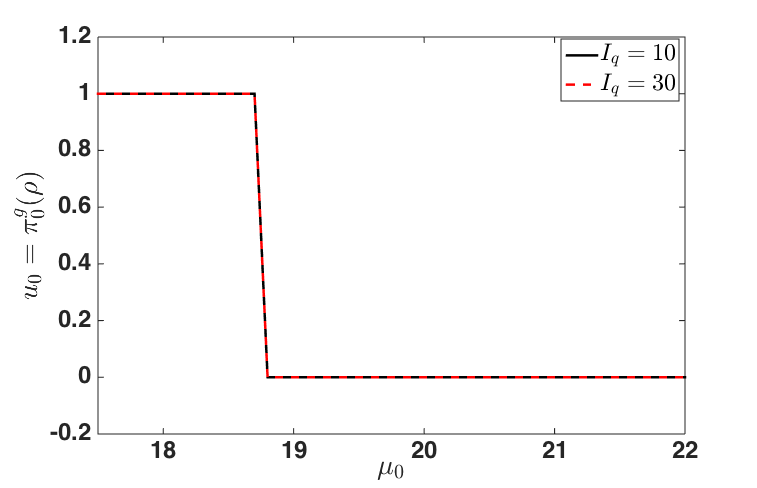}}
		\caption{\small{Comparison of optimal control inputs as a function of $\rho = \phi(x; \mu_0, 1)$ with $q_0=0$, using the finite state approximation (a) and Gaussian mixture approximation (b).  In both (a) and (b), $\delta^y = 0.5$.  Fig. (a) compares control inputs for $\delta^x = 0.1$ (black dashed line) and $\delta^x = 0.01$ (red solid line). Fig. (b) compares control inputs for $I_q = 10$ (black dashed line) and $I_q = 30$ (red solid line), which in this case are the same.  All approaches produce a thresh-hold policy that turns the heater off for $\mu_0>\approx 18.7$
				.}} 
		\label{fig:UCompare}
		\vspace{-10pt}
	\end{figure*}

\section{Numerical Example}\label{sec:example}

The temperature regulation problem is a benchmark example for hybrid systems, and a stochastic version with perfect state information is presented in \cite{Abate1}.  We consider the case of one heater, which can either be turned on to heat one room, or turned off.  The temperature of the room at time $n$ is given by the continuous variable $x_n$, and the discrete state $q_n=1$ indicates the heater is on at time $n$, and $q_n=0$ denotes the heater is off.  The stochastic difference equation governing the temperature is given by
\begin{equation*}
x_{n+1} = (1-b)x_n+cq_{n+1}+bx_a+v_n
\end{equation*}
with constants $b=0.0167$, $c=0.8$, and $x_a=6$, and $v_n$  i.i.d. Gaussian random variables with mean zero and variance $v^2$.  The control input is given by $u_n\in\mathcal{U}$ with $\mathcal{U}=\{0,1\}$, but the chosen control is not always implemented with probability $1$.  Instead, $q_n$ is updated probabilistically, dependent on $u_{n-1}$ and $q_{n-1}$, with transition function $T_q(q_{n+1}| q_n, u_n)$.  So while function $\pi_n(\sigma_n)$ deterministically returns a single control input, control input $u_n = \pi_n(\sigma_n)$ may not always be implemented.

To model this as a partially observable problem, assume the actual temperature is unknown, and only a noisy measurement is available to the controller.  The controller does, however, know whether the heater is on or off at time $n$ (i.e. $q_n$ is perfectly observed).  The observation $y_n = y^x_n$  is given by $y^x_n = x_n + w_n$, with $w_n$ i.i.d. Gaussian random variables with mean zero and variance $w^2$.

It is desirable to keep the temperature of the room between $17.5$ and $22$ degrees Celsius at all times, hence the safe region $K=[17.5,22]$ does not depend on the discrete state $q_n$ (so ${\bf 1}_K(s) = {\bf 1}_K(x)$). We consider the probability of remaining within $K$ for $N=5$ time steps given initial temperature distribution $\rho_x$ normally distributed with varying mean $\mu_0\in K$ and variance $\sP_0 =1$.  The initial mode is given as $q_0 = 0$.  The finite state and Gaussian mixture approximations are used in a PBVI algorithm in the style of Perseus \cite{Porta06}.

We consider a uniform grid ($\delta_{i,q}^x = \delta^x$ constant for all $i$, $q$) over the region $K\subset \mathbb{R}$ for the finite state approximation, with representative points at the end-point of each grid cell.  For example, setting $\delta^x=0.1$ gives $x^{1,q}=17.5, x^{2,q}=17.6,\dotsc$ for $q=0$ and $q=1$, and a total of $m_q=45$ cells $K_{i,q}$.  The function $\xi(x,q)$ maps $q$ to itself, and maps $x$ to the nearest $x^{i,q}$ in absolute value.  

The Gaussian mixture approximation utilizes an RBF approximation of the indicator function calculated using MATLAB's \emph{gmdistribution} function. We used a reduction process  to limit the number of components of each $\alpha$ and $\sigma$ for the Gaussian mixture approximation.  Similar Gaussians are combined into a single component based on the 2-norm distance between functions \cite{Zhang2010}.  Each mixture was limited to 30 components to reduce overall computation time without overly sacrificing accuracy.  This number can easily be changed, however, depending on the importance of speed versus accuracy.

Both approximations employ a sequence of sampled sets $\tilde{\Sigma}_n$ and a finite set of observations to calculate the $\alpha$-functions for the PBVI algorithm.
To generate the sets $\tilde{\Sigma}_n$, we initialized a set of 40 states $\sigma_0$ normally distributed with variance $\sP_0$ and mean $\mu_0$ randomly chosen uniformly on $K$.  Each $\sigma_0$ was updated according to $\Phi_{y,u}^g$ or $\Phi_{y,u}^{\delta}$ with $u$ chosen randomly and $y$ sampled from the corresponding $\sigma_0$ (i.e. $y\sim p(y|\sigma_0, u)$).  This process was repeated $N$ times, so that for each time step we had a set of 40 sampled $\sigma$s.  The finite set of observations were produced by a uniform grid over $\overline{K} = [16, 24]$, again using end-points as the representative observations.

To compare performance of the finite state and Gaussian mixture approximations, we present computation times and safety probability estimates for each, with varying $\delta^x$, $\delta^y$, and number of components in the indicator approximation.   Safety probabilities for varying initial distributions $\rho$ are presented in Figs. \ref{fig:discreteXcompare} and \ref{fig:gaussIndcompare} for the finite state approximation and Gaussian mixture approximation, respectively.  The optimal policy at time zero is shown for varying $\rho$ in Figs. \ref{fig:discreteXcompareU} and \ref{fig:gaussIndcompareU} for the finite and Gaussian approximations, respectively. Computation times for the Gaussian mixture approximation are given in Table \ref{table:gaussTime}, and for the finite state approximation in Table \ref{table:discreteTime}.

\begin{figure*}[th!]
	\centering
	\subfloat[]{\label{ind10}\includegraphics[width=.21\textwidth]{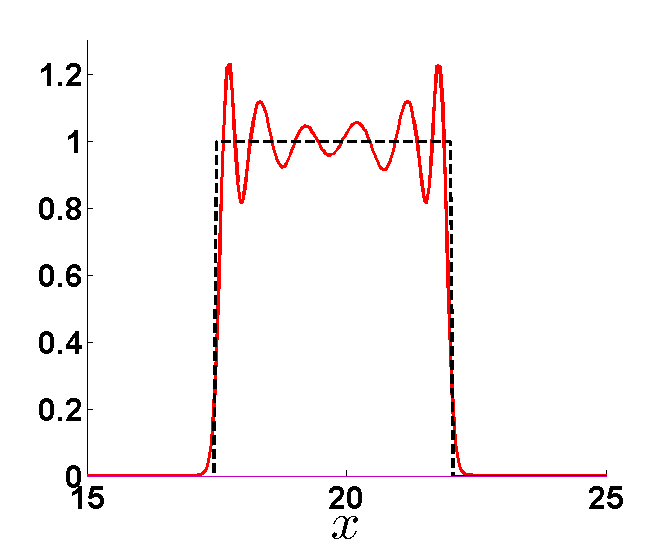}}
	\subfloat[]{\label{ind30}\includegraphics[width=.21\textwidth]{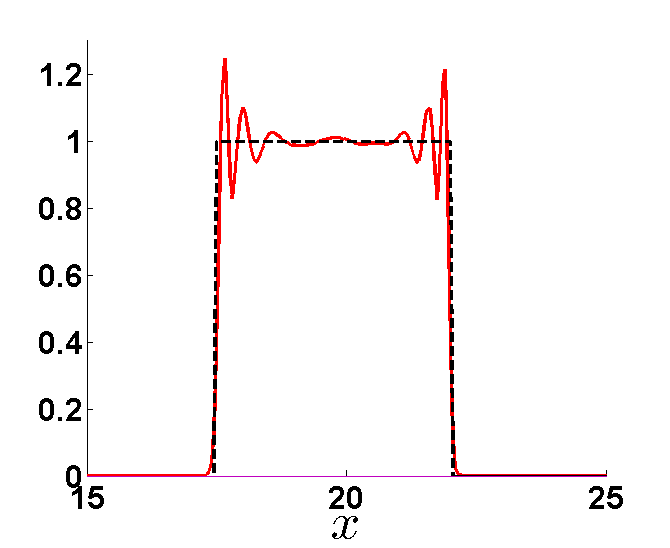}}
	\subfloat[]{\label{ind100}\includegraphics[width=.21\textwidth]{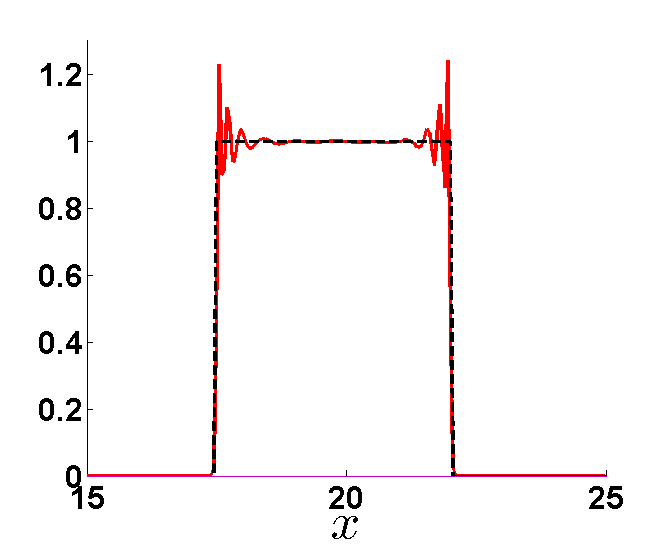}}
	\subfloat[]{\label{ind400}\includegraphics[width=.21\textwidth]{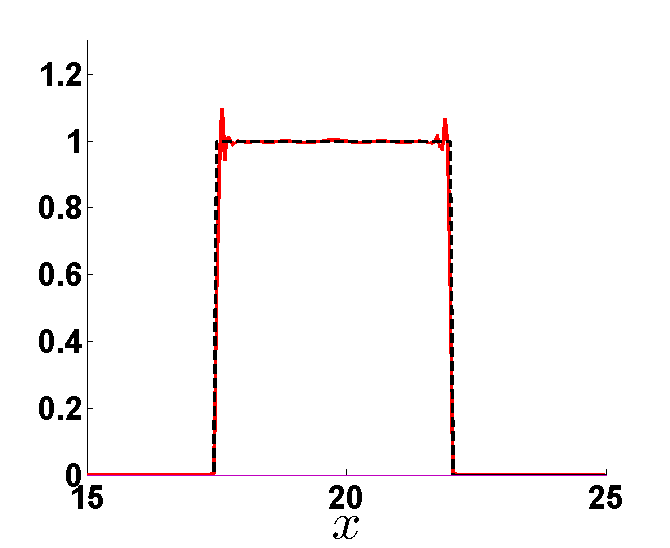}}
	\caption{\small{Comparison between ${\bf 1}_K(x)$ (in black, dashed line) to RBF approximation (red, solid line) for (a) $I_q=10$ components, (b) $I_q=30$ components, (c) $I_q=100$ components, and (d) $I_q=400$ components.  As the number of components increases, the approximation  improves, although oscillations at the endpoints remain. }}
	\label{fig:indCompare}
	\vspace{-15pt}
\end{figure*}

\begin{figure}[b!]
	\centering
	\includegraphics[width = .40\textwidth]{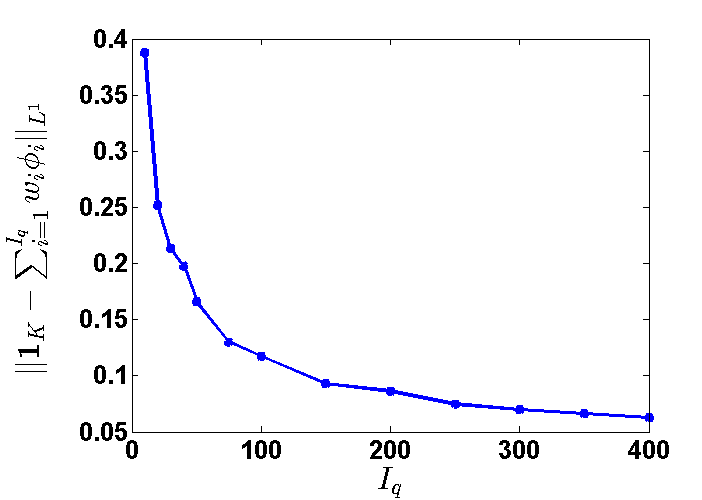}
	\caption{\small{The integrated error for RBF approximations to indicator function ${\bf 1}_K$ with a varying number of components $I_q$.  }}
	\label{fig:L1err}
\end{figure}

We also show sample RBF approximations to the indicator function ${\bf 1}_K$ in Fig. \ref{fig:indCompare} with varying numbers of components $I_q$.  The error between the RBF approximation and ${\bf 1}_K$ for varying $I_q$ is shown in Fig. \ref{fig:L1err}.  As the number of components increases, the approximation becomes more accurate, although as seen in Fig. \ref{fig:indCompare}, oscillations remain at the endpoints of $K$.  The increasing accuracy is most apparent in Fig. \ref{fig:L1err}, and demonstrates the convergence towards zero of the error $\delta^I$ with increasing $I_q$.

We show safety probabilities for  $\delta^y=0.5$.  Decreasing $\delta^y$ causes a slight increase in the safety probabilities, as expected, but there is not a significant improvement in the probability estimates, although as seen in Tables \ref{table:gaussTime} and \ref{table:discreteTime}, the increase in computation time is significant.  This is likely problem-specific, and the value of $\delta^y$ may have a greater impact for some applications.

The safety probability estimates for the finite state approximation are in general greater than for the Gaussian mixture approximation.  The mixture reduction method employed, as well as the indicator function approximation, make the Gaussian method seemingly less accurate than the finite state approximation.  However, over a finer mesh $\delta^x$, the finite state method results in greater computation time.  Note also that for $\delta^x=0.01$, the safety probabilities decrease relative to the probabilities for $\delta^x=0.1$, which highlights that although sampling from the information space and discretizing the observations guarantees a lower bound, no such guarantee exists when discretizing the state space, and there is a chance the safety estimates do not bound the true safety probabilities from below. 

Although the coarse grid produces similar results to the fine grid ($\delta^x = 0.1$ versus $\delta^x = 0.01$), in higher dimensional problems the number of grid cells  becomes prohibitive even when $\delta^x$ is large, and the Gaussian mixture approximation may be more computationally tractable.  All scenarios produce a nearly identical optimal thresh-hold policy based on the initial mean $\mu_0$, indicating that an optimal policy may be computed fairly quickly using any of the above methods.

The computation time is unfortunately still quite high in both cases, and at this time each method is likely applicable to systems with only a few modes and continuous state dimensions of no more than two or three.  It is likely, however, that computation time can be improved by using more sophisticated point-based solvers, such as \cite{SARSOP}. Further gains may be possible through adaptive gridding techniques, similar to those in \cite{SoudjaniSiam13}, to decrease the number of finite states required without sacrificing accuracy.

\begin{table}[b!]\centering
	\ra{1.3}
	\begin{tabular}{@{}crrcrr@{}}\toprule
		& \multicolumn{2}{c}{$I_q=10$} & \phantom{abc} &\multicolumn{2}{c}{$I_q=30$} \\
		\cmidrule{2-3} \cmidrule{5-6} 
		& $\delta^y=1.0$ & $\delta^y=0.5$&  & $\delta^y=1.0$ & $\delta^y=0.5$ \\ \midrule
		
		Comp. time (s)& 365.5 & 1625.9 && 1865.0 & 5586.1\\
		
		\bottomrule
	\end{tabular}
	\caption{\small{Computation times using PBVI with Gaussian mixture approximation, for varying number of components $I_q$ for RBF approximation to ${\bf 1}_K$ and discretized observation spacing $\delta^y$.}}
	\label{table:gaussTime}
\end{table}

\begin{table*}[th!]\centering
	\ra{1.3}
	\begin{tabular}{@{}rrrrcrrr@{}}\toprule
		& \multicolumn{3}{c}{$\delta^x=0.1$} & \phantom{abc} &\multicolumn{3}{c}{$\delta^x = 0.01$} \\
		\cmidrule{2-4} \cmidrule{6-8} 
		& $\delta^y=1.0$ & $\delta^y=0.5$&$\delta^y=0.1$&  & $\delta^y=1.0$ & $\delta^y=0.5$&$\delta^y = 0.1$ \\ \midrule
		
		Comp. time (s)& 50.5 & 205.1 & 1599.8&& 8961.1 & 15343.7 & 108591.3\\
		
		\bottomrule
	\end{tabular}
	\caption{\small{Computation times using PBVI with finite state approximation, for varying continuous state spacing $\delta^x$ and discretized observation spacing $\delta^y$.}}
	\label{table:discreteTime}
\end{table*}

Interestingly, increasing the number of components in the RBF approximation to the indicator function only slightly improves the safety estimates of the Gaussian mixture approximation, although the error from increasing the number of components to 30 drops significantly.  This may be caused by the mixture reduction technique, leading to a loss in the added benefit of an increased number of components when that number is again reduced.  However, although the  error with $I_q=10$ is large, we obtain safety estimates that are quite similar to the finite state approximation. This requires further investigation, but may help in decreasing computation time without losing significant accuracy by choosing $I_q$ to be small.

\section{Conclusion}\label{sec:conc}

We have presented the first numerical results for verification and controller synthesis for safety objectives, given a partially observable DTSHS.   We have considered two approximations that enable the use of a well-known POMDP optimization technique.  The first approximation discretizes the state space over a compact set $K$ and enables a vector representation of the information states and $\alpha$-functions.  The second approximates the indicator function over compact set $K$ using a finite set of Gaussian radial basis functions and enables a Gaussian mixture representation of the information states and $\alpha$-functions.  We can apply point-based value iteration to either approximation, and guarantee a lower bound to the safety probability, which is proven to converge to the true safety probability of the original PODTSHS.  A simple numerical example shows that both methods provide similar safety estimates.  The finite state approximation is faster when a coarse discretization is used, but quickly becomes slower than the Gaussian mixture approximation with a finer discretization.  Therefore,  although the Gaussian mixture produces lower safety estimates, it may be better suited to higher dimensional problems.

Although we present a switched affine system with additive Gaussian noise, both approximations may be extended to non-Gaussian systems.  Convergence results for the finite state approximation apply to arbitrary transition kernels $T_x$ and $Y_x$, given they still satisfy certain Lipschitz conditions.  The Gaussian mixture approximation further requires approximating $T_x$, $T_q$, and $Y_x$ with Gaussian mixtures, and introduces additional error.  We also focus on the safety problem, although the computational techniques presented will apply to other verification properties such as reachability, reach-avoid objectives, and others, by modifying the information state and $\alpha$-functions slightly.  We are currently working to formally extend these results to other verification objectives and more complex applications.  Because both methods are relatively slow, we plan to continue to refine them to decrease computation time, which is possible through the use of more sophisticated existing point-based solvers.  We are also exploring more efficient computation by exploiting problem structure, through the use of adaptive gridding schemes and other representations of $\alpha$ and $\sigma$ beyond vectors and Gaussian mixtures.

\appendix

\begin{IEEEproof}[Proof of Lemma \ref{lem:sigAlpha}]
		By induction. At time $N$,
		\begin{equation*}
		V_N^*(\sigma_N) =  \int_{\sS}  {\bf 1}_K(s) \sigma_N(s)\,ds.
		\end{equation*}
		By defining $\alpha_N(s) = {\bf 1}_K(s)$, 
		we obtain the desired result.  Note that this definition of $\alpha_N$ is in line with the definition given in Section \ref{sec:back_pomdp}, because although it does not represent a full policy tree (being at the terminal time, there are no more branches on the tree), it does represent the immediate value of being in state $(x,q)$, given by ${\bf 1}_K(x,q)$.
		
		Next, assuming $V_{n+1}^*(\sigma_{n+1}) = \sup_{\alpha_{n+1}} \,\langle \alpha_{n+1}, \sigma_{n+1} \rangle$, $V_n^*$ can be written as
		\begin{align*}
		V_n^*(\sigma_n) &= \max_{u\in\mathcal{U}}\mathbb{E}^{\pi}\left[\sup_{\alpha_{n+1}\in\Gamma_{n+1}}\langle \alpha_{n+1}, \sigma_n \rangle\right]\notag\\
		&= \max_{u\in\mathcal{U}}\int_{\mathcal{Y}}\sup_{\alpha_{n+1}\in\Gamma_{n+1}}\langle \alpha_{n+1}, \Phi_{y,u}\sigma_n \rangle \mathbb{P}(dy|\sigma_n, u) \\
		&= \max_{u\in\mathcal{U}}\int_{\mathcal{Y}}\sup_{\alpha_{n+1}\in\Gamma_{n+1}}\left[\int_K\int_{\sS} \alpha_{n+1}(s')\gamma(y|s') \tau(s'|s,u)\sigma_n(s)\,ds'\,ds\right]dy.\label{eq:Vn}
		\end{align*}
		Then for a specific observation $y$, action $u$, and  function $\alpha_{n+1}^i$ (index $i$ used to identify a particular $\alpha_{n+1}\in\Gamma_{n+1}$), the function $\alpha_{y,u}^i$ can be defined as
		\begin{equation}\label{alpha_yu}
		\alpha_{y,u}^i(s) = \int_{\sS}\alpha_{n+1}^{i}(s')\gamma(y|s')\tau(s'|s,u)\,ds'{\bf 1}_K(s).
		\end{equation}
		Because $\alpha_{y,u}^i$ does not depend on $\sigma_n$, we can redefine the supremum over all $\Gamma_{n+1}$ to be over all $\alpha_{y,u}^i$.
		\begin{equation*}
		V_n^*(\sigma_n) =  \max_{u\in\mathcal{U}} \int_{\sY} \sup_{\{\alpha_{y,u}^i\}} \,\langle \alpha_{y,u}^i, \sigma_n \rangle \,dy
		\end{equation*}
		For a specific $\sigma$, $u$, and $y$, we define
		\begin{equation}\label{alpha_yus}
		\begin{aligned}
		\alpha_{y,u,\sigma}(s) &= \arg\sup_i\, \langle \alpha_{y,u}^i, \sigma \rangle \\
		&= \int_{\sS}\alpha_{n+1}^{*(y)}(s')\gamma(y|s')\tau(s'|s,u)\,ds'{\bf 1}_K(s)
		\end{aligned}
		\end{equation}
		with $*(y)$ denoting the index $i$ of the $\alpha$-function $\alpha_{y,u}^i$ that maximizes the inner product. We further simplify $V_n^*$ as
		\begin{equation}\label{eq:Vint}
		\begin{aligned}
		V_n^*(\sigma_n) &=  \max_{u\in\mathcal{U}} \int_{\sY} \langle \alpha_{y,u,\sigma}, \sigma_n\rangle \,dy \\
		&=  \max_{u\in\mathcal{U}} \,\left\langle  \int_{\sY} \alpha_{y,u,\sigma}\,dy, \sigma_n \right\rangle.
		\end{aligned}
		\end{equation}
		Therefore, the set of all $\{\alpha_n\}$ is described by
		\begin{equation}\label{alpha_set_cont}
		\Gamma_n = \bigcup_{\sigma}\left\{\int_{\sY} \alpha_{y,u^*,\sigma}\,dy\right\}
		\end{equation}
		with $u^*$ the control inputs chosen according to \eqref{eq:Vint}, and $V_n^*$ may be written as 
		\begin{equation}
		V_n^*(\sigma_n) = \sup_{\alpha_n\in\Gamma_n} \,\langle \alpha_n, \sigma_n \rangle
		\end{equation}
\end{IEEEproof}

\begin{IEEEproof}[Proof of Theorem \ref{thm:sigDiscrete}]
	By induction.  At time $n=0$, $\sigma_0(s) = \rho(s) = \hat{\sigma}_0(s)$ and the inequality is trivially satisfied.  For all $i = 0,\dotsc,n$, assume $|\sigma_i(s) - \hat{\sigma}_i(s)| \leq \eta_i^{\sigma}\delta^x$.  At time $i=n+1$, for any $y\in\mathcal{Y}$ and any $u\in\mathcal{U}$,
	\begin{multline*}
	|\sigma_{n+1}(s') - \hat{\sigma}_{n+1}(s')|\leq \min\left\{\frac{1}{p(y|\sigma_n,u)}, \frac{1}{p(y|\hat{\sigma}_n, u)}\right\} \left|\gamma(y|s') \int_K\tau(s'|s,u)\sigma_n(s)\,ds \right. \\ 
	\\- \left.\gamma(y|\xi(s'))\int_K\tau(s'|\xi(s),u)\hat{\sigma}_n(s)\,ds\right|.
	\end{multline*}
	We add and subtract $\gamma(y|\xi(s'))\int_K\tau(s'|s,u)\sigma_n(s)\,ds$ and $\gamma(y|\xi(s'))\int_K\tau(s'|\xi(s),u)\sigma_n(s)\,ds$, apply the triangle inequality, and use the Lipschitz inequalities \eqref{eq:lipschitz}.
	\begin{align*}
	&|\sigma_{n+1}(s') - \hat{\sigma}_{n+1}(s')|  \leq \min\left\{\frac{1}{p(y|\sigma_n,u)}, \frac{1}{p(y|\hat{\sigma}_n, u)}\right\} \left[ \left|\gamma(y|s') - \gamma(y|\xi(s'))\right|\int_K\tau(s'|s,u)\sigma_n(s)\,ds \right.\\
	& \hspace{10 mm}\left.+ \gamma(y|\xi(s'))\int_K \left|\tau(s'|s,u) - \tau(s'|\xi(s),u)\right|\sigma_n(s)\,ds
+ \gamma(y|\xi(s'))\int_K \tau(s'|\xi(s),u)\left|\sigma_n(s) - \hat{\sigma}_n(s)\right|ds \right]\notag\\
	&\hspace{10mm}\leq \min\left\{ \frac{1}{p(y|\sigma_n,u)}, \frac{1}{p(y|\hat{\sigma}_n, u)}\right\}
	\left[\vphantom{\sum_{q\in\mathcal{Q}}\int_{K_q}}  h_y^{(2)}\|x' - \xi(x')\|\phi_v^*
 + \phi_w^*h_x^{(2)}\|x' - \xi(x')\|+ \phi_w^*h_q\|x - \xi(x)\|\phi_v^*\right. \notag\\
	&\hspace{35 mm} \left.+ \phi_w^*|\sigma_n - \hat{\sigma}_n|\sum_{q\in\mathcal{Q}}T_q(q'|s,u)\int_{K_q}\tau_x(x'|q',\xi(s),u)\,dx\right] \label{eq:thmSig2}
	\end{align*}
	Since $\tau_x$ is bounded by $\phi_v^*$, and the Lebesgue measure of $K_q$ is at most $\lambda$, we obtain
	\begin{align*}
	|\sigma_{n+1}(s') - \hat{\sigma}_{n+1}(s')|  \leq \min\left\{\frac{1}{p(y|\sigma_n,u)}, \frac{1}{p(y|\hat{\sigma}_n, u)}\right\}
	\left[\phi_v^*h_y^{(2)}\delta^x + \phi_w^*h_x^{(2)}\delta^x + \phi_w^*\phi_v^*h_q\delta^x + \phi_w^* N_q\lambda|\sigma_n - \hat{\sigma}_n|\right]. \\
	\end{align*}
	Combining terms and applying the induction hypothesis gives the desired result.
\end{IEEEproof}

\begin{IEEEproof}[Proof of Lemma \ref{lem:finiteIntY}]
	We exploit properties of the derivative of a Gaussian distribution, which bounds the Lipschitz constants for $\gamma_x$ from above. For clarity of notation, we will write $C$ rather than $C(q)$, with dependence on $q$ implicit. The constant $h_y^{(2)}$ is the maximum value of the derivative of $\phi(y^x; Cx, \mathcal{W})$ with respect to $x$:
	\begin{equation}\label{gaussDeriv}
	\left\|\frac{\partial \phi}{\partial x}\right\| \leq \frac{1}{(2\pi)^{\frac{n}{2}}|\mathcal{W}|^{\frac{1}{2}}\lambda_{1}^w}\| y^x-Cx\|\|C\|e^{-\frac{\|y^x-Cx\|^2}{2\lambda_{1}^w}}.
	\end{equation}
	Recall that $x^TA^{-1}x \geq \lambda_{\min}(A^{-1})\|x\|^2 = \frac{\|x\|^2}{\lambda_{\max}(A)}$ for $A$ a symmetric matrix.  Therefore $\exp(-\frac{1}{2}(y-Cx)^T\mathcal{W}^{-1}(y-Cx))\leq \exp(-\frac{\|y - Cx\|^2}{2\lambda_{\max}(\mathcal{W})})$.  Taking the derivative again with respect to $\|y - Cx\|$ and setting equal to zero, we see that the maximum of \eqref{gaussDeriv} occurs at $\|y - Cx\| = \sqrt{\lambda_1^w}$.

	Although $\|\frac{\partial \phi}{\partial x}\|\leq h_y^{(2)}$, we create a tighter bound for the case in which $\|y^x - Cx\|$ is greater than $\sqrt{\lambda_{1}^w}$ (for $y^x\in\mathbb{R}^l$ such that there exists $x\in K_{i,q}$ for which $\|y^x - Cx\|=\sqrt{\lambda_{1}^w}$, the upper bound $h_y^{(2)}$ is attained) using the following function. 
	\begin{equation*}\label{hyy}
	h_y(y^x) = \max_{x\in K_{i,q}} \left\{\frac{\|C\|}{(2\pi)^{\frac{n}{2}}|\mathcal{W}|^{\frac{1}{2}}\lambda_{1}^w}\| y^x - Cx\|e^{-\frac{\|y^x - Cx\|^2}{2\lambda_{1}^w}}\right\}.
	\end{equation*}
	Then,
	\begin{align}
	\int_{\mathbb{R}^{l}} |\gamma_x(y^x|x,q) - \gamma_x(y^x|\overline{x},q
	)|\,dy^x &\leq \int_{\mathbb{R}^{l}} h_y(y^x) \|x - \overline{x}\|\,dy^x \notag\\
	&\leq \delta_{i,q}^x\int_{\{y^x : \|y^x - Cx\|^2\leq \lambda_{1}^w, y^x\in\mathbb{R}^l, x\in K_{i,q}\}} h_y^{(2)}\,dy^x \notag\\
	&\hspace{10mm}+ \delta_{i,q}^x\int_{\{y^x : \|y^x - Cx\|^2> \lambda_{1}^w, y^x\in\mathbb{R}^l, x\in K_{i,q}\}} h_y(y^x)\,dy^x \notag\\
	&=\delta_{i,q}^x\beta_{1,i,q}^yh_y^{(2)}+ \delta_{i,q}^x\int_{\{y^x : \|y^x - Cx\|^2> \lambda_{1}^w, y^x\in\mathbb{R}^l, x\in K_{i,q}\}} h_y(y^x)\,dy^x\label{eq:lemAlpha22}
	\end{align}
	We use the change of variable $v = \| y^x - Cx^*\|$, with $x^* = \arg\min_{x\in K_{i,q}} \|y^x - Cx\|$, to rewrite the second term of \eqref{eq:lemAlpha22}, and apply an identity for integrals of polynomials.
	\begin{align}
	&\int_{\{y^x : \|y^x - Cx\|^2> \lambda_{1}^w, y^x\in\overline{K}, x\in K_{i,q}\}} h_y(y^x)\,dy^x\notag\\
	&\hspace{10mm}= \frac{\|C\|}{(2\pi)^{\frac{n}{2}} |\mathcal{W}|^{\frac{1}{2}}\lambda_{1}^w}\int_{\sqrt{\lambda_{1}^w}}^{\infty} ve^{-\frac{v^2}{2\lambda_{1}^w}}\,dv \notag\\
	&\hspace{10mm}\leq \frac{\|C\|}{(2\pi)^{\frac{n}{2}} |\mathcal{W}|^{\frac{1}{2}}\lambda_{1}^w}\int_{0}^{\infty} ve^{-\frac{v^2}{2\lambda_{1}^w}}\,dv \notag\\
	&\hspace{10mm}\leq \frac{\|C\|}{(2\pi)^{\frac{n}{2}} |\mathcal{W}|^{\frac{1}{2}}} \label{expIntProp}
	\end{align}
	Inserting  \eqref{expIntProp} into \eqref{eq:lemAlpha22} proves the lemma.
\end{IEEEproof}

\begin{IEEEproof}[Proof of Lemma \ref{lem:alphaTilde}]
	By induction.  At time $N$, 
	\begin{equation*}
	|\alpha_N^i(s) - \tilde{\alpha}_N^i(s)| = \left|\int_{\sS}\left({\bf 1}_K(s) - {\bf 1}_{K_{\delta}}(\xi(s))\right)ds\right| = 0
	\end{equation*}
	since for any $s\in K$, by definition $\xi(s)\in K_{\delta}$.  Assume for all $j=N-1,\dotsc,n+1$, $|\alpha_j^i(s) - \tilde{\alpha}_j^i(s)| \leq (N-n)N_q\left[\beta_1^yh_y^{(2)} + \beta_1^xh_x^{(2)} + \beta_2^y + \beta_2^x  + h_q\right]\delta^x$.  For $j=n$, separate terms and apply the triangle inequality as in the proof of Theorem \ref{thm:sigDiscrete}.
	\begin{align}
	|\alpha_n^i(s) - \tilde{\alpha}_n^i(s)| &= \left|\int_{\sS}\int_{\sY}\alpha_{n+1}^{i(y)}(s')\gamma(y|s')\tau(s'|s,u^i){\bf 1}_K(s)\,dy\,ds'\right.  \notag\\
	&\hspace{20 mm} \left.- \int_{\sS}\int_{\sY}\tilde{\alpha}_{n+1}^{i(y)}(s')\gamma(y|\xi(s'))\tau(s'|\xi(s),u^i) {\bf 1}_{K_{\delta}}(\xi(s))\,dy\,ds'\right|\notag\\
	&\leq \left|\alpha_{n+1}^{i(y)}(s') - \tilde{\alpha}_{n+1}^{i(y)}(s')\right| + N_q\left[\beta_1^yh_y^{(2)} + \beta_2^y\right]\delta^x+ N_q\left[h_q + \beta_1^xh_x^{(2)} + \beta_2^x\right]\delta^x\label{eq:alphatil2}
	\end{align}
	The second term of \eqref{eq:alphatil2} comes from Lemma \ref{lem:finiteIntY} and noting that $\alpha(s)$ represents a probability that is bounded above by one.  The third term comes from Lemma \ref{lem:finiteIntX} and the Lipschitz inequality for $T_q$ \eqref{eq:lipschitz}.  The term ${\bf 1}_K(s)$ does not affect the bound, and only indicates that both $\alpha_n(s)$ and $\tilde{\alpha}_n(s)$ are equal to zero for $s \notin K$.  Applying the induction hypothesis to \eqref{eq:alphatil2} gives the desired result.
\end{IEEEproof}

\begin{IEEEproof}[Proof of Theorem \ref{thm:ValueFuncErrorDisc}]
	By construction.  At any time $n\in[0,N]$, given $\sigma_n\in\Sigma$ and $\sigma_{n,\delta} \in\Sigma_{\delta}$, we can rewrite the value function evaluated at $\sigma$ in terms of $\alpha$-functions.
	\begin{align*}
	\left|V_n^*(\sigma_n) - V_{n,\delta}^*(\sigma_{n,\delta})\right|&= \left|\sup_{\alpha_n\in\Gamma_n}\langle \alpha_n,\sigma_n\rangle - \sup_{\alpha_{n,\delta}\in\Gamma_{n,\delta}}\langle \alpha_{n,\delta}, \sigma_{n,\delta}\rangle\right|\\
	&= \left|\langle \alpha_n^k, \sigma_n \rangle - \langle \alpha_{n,\delta}^l, \sigma_{n,\delta}\rangle \right|
	\end{align*}
	Assume without loss of generality that $\langle \alpha_n^k, \sigma_n \rangle  \geq \langle \alpha_{n,\delta}^l, \sigma_{n,\delta}\rangle$.  Then, because $\langle \tilde{\alpha}_n^k, \hat{\sigma}_n \rangle \leq \langle \hat{\alpha}_n^l, \hat{\sigma}_n \rangle$ by definition of the optimality of $\hat{\alpha}_n^l$, we can write
	\begin{align}
	\left|V_n^*(\sigma_n) - V_{n,\delta}^*(\sigma_{n,\delta})\right| &= \langle \alpha_n^k, \sigma_n \rangle - \langle \alpha_{n,\delta}^l, \sigma_{n,\delta}\rangle\notag\\
	& \leq \langle \alpha_n^k, \sigma_n \rangle - \langle \tilde{\alpha}_{n,\delta}^k, \sigma_{n,\delta}\rangle\notag\\
	& \leq \left|\langle \alpha_n^k, \sigma_n \rangle - \langle \tilde{\alpha}_n^k, \sigma_n \rangle\right|+ \left|\langle \tilde{\alpha}_n^k, \sigma_n \rangle - \langle \tilde{\alpha}_n^k, \hat{\sigma}_n \rangle\right|\notag\\
	& \leq \int_{\mathcal{S}}\left|\alpha_n^k(s) - \tilde{\alpha}_n^k(s)\right|\sigma_n(s)\,ds + \int_{\sS}\tilde{\alpha}_{n}^k(s)\left|\sigma_n(s) - \hat{\sigma}_n(s)\right|\,ds\label{eq:alphThm1}
	\end{align}
	Applying Lemma \ref{lem:alphaTilde} to the first term of \eqref{eq:alphThm1}, and noting that the integral in the second term is in fact taken over $K$ rather than $\sS$ since $\tilde{\alpha}_n^k(s)$ is zero for all $s\notin K$, we obtain
	\begin{equation}
	\left|V_n^*(\sigma_n) - V_{n,\delta}^*(\sigma_{n,\delta})\right|\leq (N-n)N_q\left[h_q + \beta_1^yh_y^{(2)} + \beta_1^xh_x^{(2)} + \beta_2^y + \beta_2^x\right]\delta^x + N_q\lambda\eta_n^{\sigma}\delta^x
	\end{equation}
	which completes the proof.
\end{IEEEproof}

\begin{IEEEproof}[Proof of Lemma \ref{lem:PhiGauss}]
The proof of both Lemma \ref{lem:PhiGauss} and Lemma \ref{lem:alphaGauss} require the following Gaussian density identities.  First, the product of two Gaussian densities is again a Gaussian density, up to a constant factor.
\begin{equation}
\label{eq:gaussprod}
\phi(x; \mu_1, \sP_1)\phi(x; \mu_2, \sP_2) = \phi(\mu_1; \mu_2, \sP_1 + \sP_2)\phi(x; \tilde{\mu}, \tilde{\sP})
\end{equation}
with
\begin{equation}
\label{eq:gaussprod2}
\begin{aligned}
\tilde{\mu} &= \tilde{\sP}\left(\sP_1^{-1}\mu_1 + \sP_2^{-1}\mu_2\right)\\
\tilde{\sP} &= \left(\sP_1^{-1} + \sP_2^{-1}\right)^{-1}
\end{aligned}
\end{equation}
Second, for invertible matrix $A$, constant $b$, and variables $x$ and $y$, 
\begin{equation}
\label{eq:gaussA}
\phi(y; Ax+b, \sP) = |A^{-1}|\phi(x; A^{-1}(y - b), A^{-1}\sP A^{-T}).
\end{equation}
Both identities are easily shown directly.

The proof then follows by construction.  Given $\sigma_{n,g}(x,q) =\displaystyle \sum_{l=1}^L w^{\sigma}_{l,n}(q)\phi(x; \mu^{\sigma}_{l,n}(q), \sP^{\sigma}_{l,n}(q))$, observation $y\in\sY$, and control input $u\in\sU$, the operator $\Phi_{y,u}^g\sigma_{n,g}$ produces
\begin{align}\label{eq:phiG}
\Phi_{y,u}^g\sigma_{n,g} &= \gamma(y|s',u)\sum_{q\in\mathcal{Q}}\int_{\mathbb{R}^m}\left[\sum_{i=1}^{I_q} w_i^I(q) \phi_i^I(x)\right]\tau(s'|s,u)\sigma_{n,g}(s)\,dx.
\end{align}
  We use the notation $w_i^I$, $\mu_i^I$, and $\sP_i^I$ for the weights, means, and covariances of the indicator function approximation to distinguish them from the other Gaussian mixtures representing $\sigma$ and $\alpha$.  Replacing $\sigma_{n,g}$ by its Gaussian mixture representation in \eqref{eq:phiG}, and expanding $\gamma$ and $\tau$, gives
\begin{align}
\left(\Phi_{y,u}^g\sigma_{n,g}\right)(Cx',q') &= Y_{q}(y^q|q')\phi(y^x; Cx',\sW)\sum_{q\in\sQ}\int_{\mathbb{R}^m}\left[\sum_{i=1}^{I_q} w_i^I(q) \phi_i^I(x)\right] T_q(q'|q,u)\phi(x';Ax+g(q',u),\sV) \notag\\
&\hspace{30 mm}\times\left[\sum_{l=1}^L w^{\sigma}_{l,n}(q)\phi(x; \mu^{\sigma}_{l,n}(q), \sP^{\sigma}_{l,n}(q))\right]\\
 &= \sum_{q=1}^{N_q}\sum_{l=1}^L\sum_{i=1}^{I_q}w_i^I(q)w_{m,n}^{\sigma}(q)Y_{q}(y^q|q')T_q(q'|q,u)|C^{-1}|\phi(x';C^{-1}y^x; C^{-1}\sW C^{-T})\notag\\
&\hspace{20 mm}\times \int_{\mathbb{R}^m}\phi(x; \mu_i^I(q), \sP_i^I(q))\phi(x; \mu_{l,n}^{\sigma}(q), \sP_{l,n}^{\sigma}(q))\notag\\
&\hspace{30 mm}\times|A^{-1}|\phi(x; A^{-1}(x'-g(q',u)), A^{-1}\sV A^{-T})\,dx\label{eq:phi1}\\
&=\sum_{q=1}^{N_q}\sum_{l=1}^L\sum_{i=1}^{I_q}|A^{-1}||C^{-1}|w_i^I(q)w_{m,n}^{\sigma}(q)Y_{q}(y^q|q')T_q(q'|q,u)\phi(x';C^{-1}y^x; C^{-1}\sW C^{-T})\notag\\
&\hspace{20 mm}\times\phi(\mu_i^I(q); \mu_{l,n}^{\sigma}(q), \sP_i^I(q) + \sP_{l,n}^{\sigma}(q))\int_{\mathbb{R}^m}\phi(x; \tilde{\mu},\tilde{\sP})\notag\\
&\hspace{30 mm}\times \phi(x; A^{-1}(x'-g(q',u)), A^{-1}\sV A^{-T})\,dx \label{eq:phi2}\\
&=\sum_{q=1}^{N_q}\sum_{l=1}^L\sum_{i=1}^{I_q}|A^{-1}||C^{-1}|w_i^I(q)w_{m,n}^{\sigma}(q)Y_{q}(y^q|q')T_q(q'|q,u)\phi(x';C^{-1}y^x, C^{-1}\sW C^{-T})\notag\\
&\hspace{20 mm}\times \phi(\mu_i^I(q); \mu_{l,n}^{\sigma}(q), \sP_i^I(q) + \sP_{l,n}^{\sigma}(q))\phi(A^{-1}(x'-g(q',u)); \tilde{\mu}, \tilde{\sP} + A^{-1}\sV A^{-T})\label{eq:phi3}
\end{align}
Line \eqref{eq:phi1} follows from \eqref{eq:gaussA}, line \eqref{eq:phi2} from combining $\phi(x; \mu_i^I(q), \Sigma_i^I(q))$ and $\phi(x; \mu_{l,n}^{\sigma}(q), \sP_{l,n}^{\sigma}(q))$ according to \eqref{eq:gaussprod}, and \eqref{eq:phi3} from a second application of \eqref{eq:gaussprod} and setting the integral of a Gaussian density over $\sX$ equal to one.

A final application of \eqref{eq:gaussprod} and \eqref{eq:gaussA} gives
\begin{align*}
\left(\Phi_{y,u}^g\sigma_{n,g}\right)(x',q') &=\sum_{q=1}^{N_q}\sum_{l=1}^L\sum_{i=1}^{I_q}w_i^I(q)w_{m,n}^{\sigma}(q)Y_{q}(y^q|q')T_q(q'|q,u) \phi(\mu_i^I(q); \mu_{l,n}^{\sigma}(q), \sP_i^I(q) + \sP_{l,n}^{\sigma}(q))\notag\\
&\hspace{10 mm}\times\phi(y^x; C(A\tilde{\mu}+g(q',u)),\sW+C\sV C^T+CA\tilde{\sP}A^TC^T)\phi(x; \mu_{q,l,i,n+1}^{\sigma}(q'), \sP_{q,l,i,n+1}^{\sigma}(q'))
\end{align*}
which is again a Gaussian mixture with $N_qLI_q$ components.  Specifically, 
\begin{equation*}
\sigma_{n+1,g}(x',q') = \sum_{q=1}^{N_q}\sum_{l=1}^L\sum_{i=1}^{I_q}w_{q,l,i,n+1}^{\sigma}(q')\phi(x';\mu_{q,l,i,n+1}^{\sigma}(q'), \sP_{q,l,i,n+1}^{\sigma}(q')),
\end{equation*}
with
\begin{align*}
w_{q,l,i,n+1}^{\sigma}(q') &= w_i^I(q)w_{l,n}^{\sigma}(q)Y_{q}(y^q|q')T_q(q'|q,u) \phi(\mu_i^I(q); \mu_{l,n}^{\sigma}(q),\sP_{i}^I(q) + \sP_{l,n}^{\sigma}(q))\\
&\hspace{20 mm}\times\phi(y^x; C(A\tilde{\mu}+g(q',u)), \sW + C\sV C^T + C A\tilde{\sP}A^TC^T), \\
\mu_{q,l,i,n+1}^{\sigma}(q') &=\sP_{q,l,i,n+1}^{\sigma}(q')\left[C^T\sW^{-1}y^x + \left(A\tilde{\sP}A^T + \sV\right)^{-1}\left(A\tilde{\mu} + g(q,u,q')\right)\right],\\
\sP_{q,l,i,n+1}^{\sigma}(q') &= \left[C^T\sW^{-1}C + \left(A\tilde{\sP}A^T+\sV\right)^{-1}\right]^{-1},
\end{align*}
and
\begin{align*}
\tilde{\mu} &= \tilde{\sP}\left[\left(\sP_i^I(q)\right)^{-1}\mu_i^I(q) + \left(\sP_{l,n}^{\sigma}(q)\right)^{-1}\mu_{l,n}^{\sigma}(q)\right],\\
\tilde{\sP} &= \left[\left(\sP_i^I(q)\right)^{-1} + \left(\sP_{l,n}^{\sigma}(q)\right)^{-1}\right].
\end{align*}

\end{IEEEproof}

\begin{IEEEproof}[Proof of Lemma \ref{lem:alphaGauss}]
	By construction.  Given $\alpha_{n+1,g}^{*(y)}(x',q') =\displaystyle\sum_{m=1}^M w_{m,n+1}^{\alpha,y}(q')\phi(x';\mu_{m,n+1}^{\alpha,y}(q'), \sP_{m,n+1}^{\alpha,y}(q'))$ for some observation $y\in\sY$, then $\alpha_{y,u,\sigma}^g(x,q)$ for control input $u\in\sU$ is written 
	\begin{align}
	\alpha_{y,u,\sigma}^g(x,q)&=\sum_{q'\in\mathcal{Q}}\int_{\mathbb{R}^m} \alpha_{n+1,g}^{*(y)}(s')\gamma(y|s')\tau(s'|s,u)\,dx'
	\left[\sum_{i=1}^{I_{q}} w_i^I(q) \phi_i^I(x) \right]\label{eq:alph}\\
	&= \sum_{q'\in\mathcal{Q}}\int_{\mathbb{R}^m}\left[\sum_{m=1}^M w_{m,n+1}^{\alpha,y}(q')\phi(x';\mu_{m,n+1}^{\alpha,y}(q'), \sP_{m,n+1}^{\alpha,y}(q'))\right]\notag\\
	&\hspace{20 mm}\times Y_q(y^q|q')\phi(y^x; Cx',\sW)T_q(q'|q,u)\notag \\
	&\hspace{30 mm}\times\phi(x';Ax+g(q',u),\sV)\,dx'  \left[\sum_{i=1}^{I_{q}} w_i^I(q) \phi_i^I(x) \right]\label{eq:alph1}\\
	&=\sum_{q'=1}^{N_q}\sum_{m=1}^M\sum_{i=1}^{I_q}w_{m,n+1}^{\alpha,y}(q')w_i^I(q)Y_q(y^q|q')T_q(q'|q,u)\notag\\
	&\hspace{20 mm}\times \phi(x;\mu_i^I(q),\sP_i^I(q))|C^{-1}|\phi(C^{-1}y^x; \mu_{m,n+1}^{\alpha,y}(q'), C^{-1}\sW C^{-T} + \sP_{m,n+1}^{\alpha,y}(q'))\notag\\
	&\hspace{30 mm}\times\int_{\mathbb{R}^m}\phi(x';\hat{\mu}, \hat{\sP})\phi(x';Ax+g(q',u),\sV)\,dx'\label{eq:alph2}\\
	&=\sum_{q'=1}^{N_q}\sum_{m=1}^M\sum_{i=1}^{I_q}w_{m,n+1}^{\alpha,y}(q')w_i^I(q)Y_q(y^q|q')T_q(q'|q,u)\notag\\
	&\hspace{20 mm}\times \phi(x;\mu_i^I(q),\sP_i^I(q))\phi(y^x; C\mu_{m,n+1}^{\alpha,y}(q'), \sW + C\sP_{m,n+1}^{\alpha,y}(q')C^T)\notag\\
	&\hspace{30 mm}\times |A^{-1}|\phi(x; A^{-1}(\hat{\mu}-g(q',u)), A^{-1}(\hat{\sP}+\sV)A^{-T})\label{eq:alph3}
	\end{align}
	\begin{align}
	&=\sum_{q'=1}^{N_q}\sum_{m=1}^M\sum_{i=1}^{I_q}w_{m,n+1}^{\alpha,y}(q')w_i^I(q)Y_q(y^q|q')T_q(q'|q,u)\notag\\
	&\hspace{20 mm}\times \phi(\mu_i^I(q); A^{-1}(\hat{\mu}-g(q',u)), \sP_i^I(q)+A^{-1}(\hat{\sP}+\sV)A^{-T})\notag\\
	&\hspace{30 mm}\times \phi(y^x; C\mu_{m,n+1}^{\alpha,y}(q'), \sW + C\sP_{m,n+1}^{\alpha,y}(q')C^T)\phi(x; \mu_{q',m,i,n}^{\alpha,y}(q),\sP_{q',m,i,n}^{\alpha,y}(q))\label{eq:alph4}
	\end{align}
	Line \eqref{eq:alph2} follows from one application of \eqref{eq:gaussprod}, line \eqref{eq:alph3} from \eqref{eq:gaussA} and another application of \eqref{eq:gaussprod}, and a final product of Gaussian densities gives \eqref{eq:alph4}.  Hence $\alpha_{y,u,\sigma}^g$ is a Gaussian mixture with $N_qMI_q$ components.  
	
	Recalling that $\alpha_{n,g}(x,q) = \sum_{y\in\sY}\alpha_{y,u,\sigma}^g$, it follows that 
\begin{equation*}
\alpha_{n,g}(x,q) = 
\sum_{y\in\sY}\sum_{q=1}^{N_q}\sum_{m=1}^M\sum_{i=1}^{I_q}w_{y,q',m,i,n+1}^{\sigma}(q)\phi(x;\mu_{y,q',m,i,n+1}^{\alpha}(q), \sP_{y,q',m,i,n+1}^{\alpha}(q)),
\end{equation*}
with
\begin{align*}
w_{y,q,l,i,n+1}^{\alpha}(q)&=w_{m,n+1}^{\alpha,y}(q')w_i^I(q)Y_q(y^q|q')T_q(q'|q,u)\notag\\
&\hspace{10 mm}\times \phi(\mu_i^I(q); A^{-1}(\hat{\mu}-g(q',u)), \sP_i^I(q)+A^{-1}(\hat{\sP}+\sV)A^{-T})\notag\\
&\hspace{20 mm}\times \phi(y^x; C\mu_{m,n+1}^{\alpha,y}(q'), \sW + C\sP_{m,n+1}^{\alpha,y}(q')C^T)\notag\\
\mu_{y,q',m,i,n+1}^{\alpha}(q) &= \sP_{y,q',m,i,n+1}^{\alpha}(q)\left[\left(\sP_i^I(q)\right)^{-1}\mu_i^I(q) + A^T\left(\hat{\sP} + \sV\right)^{-1}(\hat{\mu} - g(q',u))\right]\\
\sP_{y,q',m,i,n+1}^{\alpha}(q) &= \left[\left(\sP_i^I(q)\right)^{-1} + A^T\left(\hat{\sP}+\sV\right)^{-1}A\right]^{-1},
\end{align*}
and
\begin{align*}
\hat{\mu} &= \hat{\sP}\left[C^T\sW^{-1}y^x + \left(\sP_{m,n+1}^{\alpha,y}(q')\right)^{-1}\mu_{m,n+1}^{\alpha,y}(q')\right]\\
\hat{\sP} &= \left[C^T\sW^{-1}C + \left(\sP_{m,n+1}^{\alpha,y}(q')\right)^{-1}\right]^{-1}.
\end{align*}

\end{IEEEproof}

\begin{IEEEproof}[Proof of Theorem \ref{thm:sigContError}]
	By induction.  At time zero, $\sigma_{0,g}(s) = \sigma_0(s)$, so that $\|\sigma_{0} - \sigma_{0,g}\|_1 = 0$.  Assume that $\|\sigma_i- \sigma_{i,g}\|_1 \leq \gamma^{\sigma}_iN_q\delta^I$ for all $i=1,\dotsc,n$.  Then at time $n+1$ we have, for some $y\in\mathcal{Y}$ and $u\in\mathcal{U}$,
	\begin{align}
	\left\|\sigma_{n+1} - \sigma_{n+1,g}\right\|_1 &\leq\int_{\mathcal{S}} \gamma(y|s')\int_{\mathcal{S}}\left|{\bf 1}_K(s)\sigma_n(s)\,ds - \sum_{i=1}^{I_q}w_i(q)\phi_i(x) \sigma_{n,g}(s)\,ds\right|\tau(s'|s,u)\,ds'\notag\\
	&\leq \phi_w^*\left[\vphantom{\sum_{i=1}^{I_q}}\int_{\mathcal{S}}\left|{\bf 1}_K(s)\sigma_n(s)\,ds - {\bf 1}_K(s)\sigma_{n,g}(s)\,ds \right|\right.\notag \\
	&\hspace{10 mm}+\left. \int_{\mathcal{S}}\left|{\bf 1}_K(s)\sigma_{n,g}(s)\,ds - \sum_{i=1}^{I_q}w_i(q)\phi_i(x) \sigma_{n,g}(s)\,ds\right| \right]\notag\\
	&\leq \phi_w^*\left[ \vphantom{\sum_{i=1}^{I_q}}\left\|\sigma_n - \sigma_{n,g}\right\|_1+ \sum_{q\in\sQ}\int_{\mathbb{R}^m}\left|{\bf 1}_{K_q}(x) -  \sum_{i=1}^{I_q}w_i(q)\phi_i(x)\right|\sigma_{n,g}(x,q)\,dx\right]\label{eq:thmContError1}
	\end{align}
	The first term of line \eqref{eq:thmContError1} follows because the integral over $K$ is less than the integral over all of $\mathcal{S}$, since $K$ is a compact subset of $\mathcal{S}$.  The induction hypothesis completes the proof.
\end{IEEEproof}

\begin{IEEEproof}[Proof of Lemma \ref{lem:alphGtil}]
	By induction.  At time $N$,
	\begin{align*}
	\left\|\alpha_N - \tilde{\alpha}_{N,g}\right\|_{1} &= \sum_{q\in\sQ}\int_{\mathbb{R}^m}\left|{\bf 1}_{K_q}(x) - \sum_{i=1}^{I_q}w_i(q)\phi_i(x)\right|\,dx \leq N_q\delta^I
	\end{align*}
	and the result is satisfied.  Assume for $j=N-1,\dotsc,n+1$ that $\|\alpha_j^i - \tilde{\alpha}_{j,g}^i\|_1 \leq \left(\sum_{k=j}^N(\lambda\phi_v^*)^{N-k}\overline{\alpha}_{N-k+j+1}\right)N_q\delta^I$ for any $\alpha_j^i\in \Gamma_j$, letting $\overline{\alpha}_{N+1}=1$. Then for $j=n$,
	\begin{align*}
	\left\|\alpha_n^i(s) - \tilde{\alpha}_{n,g}^i(s)\right\|_1 &= \int_{\sS}\left|\vphantom{\sum_{i=1}^{I_q}w_i(q)\phi_i(x)}\int_{\sS}\int_{\sY}\alpha_{n+1}^{i(y)}(s')\gamma(y|s')\tau(s'|s,u^i){\bf 1}_K(s)\,dy\,ds'\right. \\
	&\hspace{20 mm} -  \tilde{\alpha}_{n+1,g}^{i(y)}(s')\gamma(y|s')\tau(s'|s,u^i)\left.\sum_{i=1}^{I_q}w_i(q)\phi_i(x)\right|\,dy\,ds'\,ds \\
	&\leq\int_{\sS}\int_{\sS}\int_{\sY}\left|\alpha_{n+1}^{i(y)}(s') - \tilde{\alpha}_{n+1,g}^{i(y)}(s')\right|\gamma(y|s')\tau(s'|s,u^i){\bf 1}_K(s)\,dy\,ds'\,ds \notag \\
	&\hspace{20 mm}+\int_{\sS}\int_{\sS}\int_{\sY}\tilde{\alpha}_{n+1,g}^{i(y)}(s')\gamma(y|s')\tau(s'|s,u^i)\left|{\bf 1}_K(s) - \sum_{i=1}^{I_q}w_i(q)\phi_i(x)\right|\,dy\,ds'\,ds\\
	&\leq \int_{\sS}\left|\alpha_{n+1}^{i(y)}(s') - \tilde{\alpha}_{n+1,g}^{i(y)}(s')\right|\phi_v^*ds'\int_{\sS}{\bf 1}_K(s)ds + \sum_{q\in\sQ}\int_{\mathbb{R}^m}\overline{\alpha}_{n+1}\left|{\bf 1}_{K_q}(x) - \sum_{i=1}^{I_q}w_i(q)\phi_i(x)\right|dx\\
	&\leq \left(\sum_{k=n}^N(\lambda\phi_v^*)^{N-k}\overline{\alpha}_{N-k+n+1}\right)N_q\delta^I 
	\end{align*}
\end{IEEEproof}

\begin{IEEEproof}[Proof of Theorem \ref{thm:ValueFuncErrorGauss}]
	By construction.  For any time $n\in[0,N]$, given $\sigma_n\in\Sigma_n$ and $\sigma_{n,g}\in\Sigma_{n,g}$, we can rewrite the value function evaluated at $\sigma$ in terms of $\alpha$-functions.
	\begin{align*}
	\left|V_n^*(\sigma_n) - V_{n,g}^*(\sigma_{n,g})\right| 
	&= \left|\sup_{\alpha_n\in\Gamma_n}\langle 
	\alpha_n, \sigma_n\rangle - \sup_{\alpha_{n,g}\in\Gamma_{n,g}}\langle \alpha_{n,g},\sigma_{n,g}\rangle\right|\\
	&= \left|\langle \alpha_n^k, \sigma_n\rangle - \langle \alpha_{n,g}^l,\sigma_{n,g}\rangle\right|
	\end{align*}
	As in the proof of Theorem \ref{thm:ValueFuncErrorDisc}, assume without loss of generality that $\langle \alpha_n^k, \sigma_n\rangle \geq \langle \alpha_{n,g}^l,\sigma_{n,g}\rangle$. 
	\begin{align*}
	\left|V_n^*(\sigma_n) - V_{n,g}^*(\sigma_{n,g})\right|&=\langle \alpha_n^k, \sigma_n\rangle - \langle \alpha_{n,g}^l,\sigma_{n,g}\rangle\\
	&\leq \langle \alpha_n^k, \sigma_n\rangle - \langle \tilde{\alpha}_{n,g}^k,\sigma_{n,g}\rangle\\
	& \leq \left|\langle \alpha_n^k, \sigma_n \rangle - \langle \tilde{\alpha}_{n,g}^k, \sigma_n \rangle \right| + \left|\langle \tilde{\alpha}_{n,g}^k,\sigma_{n}\rangle - \langle \tilde{\alpha}_{n,g}^k, \sigma_{n,g}\rangle\right|\\
	&\leq \int_{\sS}\left|\alpha_n^k(s) - \tilde{\alpha}_{n,g}^k(s)\right|\sigma_n(s)\,ds + \int_{\sS}\tilde{\alpha}_{n,g}^k(s)\left|\sigma_n(s) - \sigma_{n,g}(s)\right|\,ds\\
	&\leq \left(\sum_{k=n+1}^N(\lambda\phi_v^*)^{N-k}\overline{\alpha}_{N-k+n+1}\right)\phi_{\sigma,n}^*N_q\delta^I + \overline{\alpha}_n\gamma_n^{\sigma}N_q\delta^I
	\end{align*}
\end{IEEEproof}

\begin{IEEEproof}[Proof of Lemma \ref{lem:discObsAlpha}]
	Define $\overline{K}$ such that $Y_x(\mathcal{Y}\backslash \overline{K}|z\in K_{\delta},u)<\frac{\epsilon}{N}$. Then
	\begin{align}
	V_{n,\delta}^*(\sigma_{n,\delta}) - \hat{V}_{n,\delta}^*(\sigma_{n,\delta})  &= \sup_{\alpha_{n,\delta}\in\Gamma_{n,\delta}}\sum_{z\in K_{\delta}} \alpha_{n,\delta}(z) \sigma_{n,\delta}(z) - \sup_{\hat{\alpha}\in\tilde{\Gamma}_{n,\delta}}\sum_z\hat{\alpha}_{n,\delta}(z)\sigma_{n,\delta}(z)\notag\\
	&\leq \int_{\overline{K}}\sum_{z, z'\in K_{\delta}}\left[\alpha_{n+1,\delta}^{*(y)}(z')\gamma(y|z')\tau_{\delta}(z'|z,u^*)\sigma_{n,\delta}(z)\,dy\right.\notag\\
	&\hspace{10 mm}\left. - \alpha_{n+1,\delta}^{*(\theta(y))}(z')\gamma(y|z')\tau_{\delta}(z'|z,u^*)\sigma_{n,\delta}(z)\,dy \right] \notag\\
	&\hspace{10 mm} + \int_{\mathcal{Y}\backslash\overline{K}}\sum_{z, z'\in K_{\delta}}\alpha_{n+1,\delta}^{*(y)}(z')\gamma(y|z')\tau_{\delta}(z'|z,u^*)\sigma_{n,\delta}(z)\,dy \notag\\
	&\hspace{10 mm}- \int_{\mathcal{Y}\backslash\overline{K}}\sum_{z,z'\in K_{\delta}}\alpha_{n+1,\delta}^{*(\psi_y)}(z')\gamma(y|z') \tau_{\delta}(z'|z,u^*)\sigma_{n,\delta}(z)\,dy\notag\\
	&\leq \int_{\overline{K}}\sum_{z,z'\in K_{\delta}}\left[\alpha_{n+1,\delta}^{*(y)}(z')\gamma(y|z')\tau_{\delta}(z'|z,u^*)\sigma_{n,\delta}(z)\,dy\right.\notag\\
	&\hspace{10 mm}\left. - \alpha_{n+1,\delta}^{*(\theta(y))}(z')\gamma(y|z')\tau_{\delta}(z'|z,u^*)\sigma_{n,\delta}(z)\,dy \right]  +\frac{\epsilon}{N} \label{eq:lemdiscObs2}
	\end{align}
	Note that \eqref{eq:lemdiscObs2} is nonnegative, meaning that using $\hat{\alpha}_{\delta}^*$ produces a lower bound to the actual value function given by $\alpha_{\delta}^*$.  This follows because $\alpha_{n+1,\delta}^{*(y)}$ is chosen optimally for only a subset of $\mathcal{Y}$ (at the points $\theta(y)$), and for all other $y\in\mathcal{Y}$, $\alpha_{n+1,\delta}$ is suboptimal, producing a lower value.  
	
	Next, we can bound $\alpha_{n+1,\delta}^{*(\theta(y))}\gamma(y|z')$ from below based on how the points $w$ are defined.
	\begin{align*}
	V_{n,\delta}^*(\sigma_{\delta}) - \hat{V}_{n,\delta}^*(\sigma_{n,\delta})&\leq  \int_{\overline{K}} \left[\alpha_{n+1,\delta}^{*(y)}(z')\gamma(y|z') - \alpha_{n+1,\delta}^{*(\theta(y))}(z')\gamma(\theta(y)|z')\right]\,dy + \frac{\epsilon}{N}\\
	&\leq N_q\overline{\lambda}h_y^{(1)}\delta^y + \frac{\epsilon}{N}
	\end{align*}
\end{IEEEproof}

\begin{IEEEproof}[Proof of Theorem \ref{thm:discObsVF}]
	By induction.    At time $N$, $V_{N,\delta}^*(\sigma_{n,\delta}) = \tilde{V}_{N,\delta}^*(\sigma_{n,\delta})$ since $\Gamma_N = \tilde{\Gamma}_N = {\bf 1}_{K_{\delta}}$.  Assume for all $i=n+1,\dotsc,N-1$ that $V_{i,\delta}^*(\sigma_{n,\delta})- \tilde{V}^*_{i,\delta}(\sigma_{n,\delta}) \leq (N-i)N_q\overline{\lambda}h_y^{(1)}\delta^y + \frac{(N-j)\epsilon}{N}$.  Then, at time $n$,
	\begin{align*}
	V_{n,\delta}^*(\sigma_{n,\delta})- \tilde{V}^*_{n,\delta}(\sigma_{n,\delta})&=\langle \alpha_{n,\delta}^*, \sigma_{n,\delta}\rangle - \langle \tilde{\alpha}_{n,\delta}^*, \sigma_{n,\delta} \rangle \\
	&= \langle \alpha_{n,\delta}^*, \sigma_{n,\delta}\rangle - \langle \hat{\alpha}_{n,\delta}^*, \sigma_{n,\delta}\rangle + \langle \hat{\alpha}_{n,\delta}^*, \sigma_{n,\delta}\rangle - \langle \tilde{\alpha}_{n,\delta}^*, \sigma_{n,\delta} \rangle \\
	&\leq N_q\overline{\lambda}h_y^{(1)}\delta^y + \int_{\mathcal{Y}}\sup_{\Gamma_{n+1,\delta}}\langle \alpha_{n+1,\delta}, \Phi_{y,\hat{u}^*}^{\delta}\sigma_{n,\delta} \rangle \mathbb{P}_{
		\delta}(dy|\sigma_{n,\delta},\hat{u}^*)\notag \\
	&\hspace{15 mm}- \int_{\mathcal{Y}}\sup_{\tilde{\Gamma}_{n+1,\delta}}\langle \tilde{\alpha}_{n+1,\delta}, \Phi^{\delta}_{y,\tilde{u}^*}\sigma_{\delta} \rangle \mathbb{P}_{\delta}(dy|\sigma_{n,\delta},\tilde{u}^*)+\frac{\epsilon}{N}  \\
	&\leq  N_q\overline{\lambda}h_y^{(1)}\delta^y  + V_{n+1,\delta}^*(\sigma_{n+1,\delta})- \tilde{V}^*_{n+1,\delta}(\sigma_{n+1,\delta})+\frac{\epsilon}{N}\\
	\end{align*}
	Applying the induction inequality and combining terms completes the proof.
\end{IEEEproof}

\bibliographystyle{IEEEtran}

\begin{thebibliography}{10}
	\providecommand{\url}[1]{#1}
	\csname url@samestyle\endcsname
	\providecommand{\newblock}{\relax}
	\providecommand{\bibinfo}[2]{#2}
	\providecommand{\BIBentrySTDinterwordspacing}{\spaceskip=0pt\relax}
	\providecommand{\BIBentryALTinterwordstretchfactor}{4}
	\providecommand{\BIBentryALTinterwordspacing}{\spaceskip=\fontdimen2\font plus
		\BIBentryALTinterwordstretchfactor\fontdimen3\font minus
		\fontdimen4\font\relax}
	\providecommand{\BIBforeignlanguage}[2]{{%
			\expandafter\ifx\csname l@#1\endcsname\relax
			\typeout{** WARNING: IEEEtran.bst: No hyphenation pattern has been}%
			\typeout{** loaded for the language `#1'. Using the pattern for}%
			\typeout{** the default language instead.}%
			\else
			\language=\csname l@#1\endcsname
			\fi
			#2}}
	\providecommand{\BIBdecl}{\relax}
	\BIBdecl
	
	\bibitem{oishi1}
	C.~Tomlin, I.~Mitchell, A.~Bayen, and M.~Oishi, ``Computational techniques for
	the verification and control of hybrid systems,'' in \emph{Proceedings of the
		{IEEE}}, vol.~91, no.~7, July 2003, pp. 986--1001.
	
	\bibitem{Prand1}
	M.~Prandini and J.~Hu, \emph{Stochastic Reachability: Theoretical Foundations
		and Numerical Approximation}, ser. Lecture Notes in Control and Information
	Sciences.\hskip 1em plus 0.5em minus 0.4em\relax Springer Verlag, 2006, pp.
	107--139.
	
	\bibitem{Mitchell1}
	I.~Mitchell and J.~Templeton, ``A toolbox of {H}amilton-{J}acobi solvers for
	analysis of nondeterministic continuous and hybrid systems,'' in \emph{Hybrid
		Systems: Computation and Control}, 2005, vol. 3414, pp. 480--494.
	
	\bibitem{Abate1}
	A.~Abate, M.~Prandini, J.~Lygeros, and S.~Sastry, ``Probabilistic reachability
	and safety for controlled discrete time stochastic hybrid systems,''
	\emph{Automatica}, vol.~44, no.~11, pp. 2724--2734, 2008.
	
	\bibitem{summers}
	S.~Summers and J.~Lygeros, ``Verification of discrete time stochastic hybrid
	systems: A stochastic reach-avoid decision problem,'' \emph{Automatica},
	vol.~46, no.~12, pp. 1951--1961, 2010.
	
	\bibitem{Verma2012}
	R.~Verma and D.~del Vecchio, ``Safety control of hidden mode hybrid systems,''
	\emph{{IEEE} Transactions on Automatic Control}, vol.~57, no.~1, pp. 62--77,
	2012.
	
	\bibitem{Ghaemi2013}
	R.~Ghaemi and D.~del Vecchio, ``Control for safety specifications of systems
	with imperfect information on a partial order,'' \emph{{IEEE} Transactions on
		Automatic Control}, vol.~59, no.~4, pp. 982 -- 995, 2014.
	
	\bibitem{Ding2013}
	J.~Ding, A.~Abate, and C.~Tomlin, ``Optimal control of partially observable
	discrete time stochastic hybrid systems for safety specifications,'' in
	\emph{American Control Conference}, 2013, pp. 6231--6236.
	
	\bibitem{Lesser2014}
	K.~Lesser and M.~Oishi, ``Reachability for partially observable discrete time
	stochastic hybrid systems,'' \emph{Automatica}, vol.~50, no.~8, pp.
	1989--1998, 2014.
	
	\bibitem{Lesser2015_HSCC}
	------, ``Finite state approximation for verification of partially observable
	stochastic hybrid systems,'' in \emph{Hybrid Systems: Computation and
		Control}, 2015.
	
	\bibitem{bertsekas2}
	D.~P. Bertsekas, \emph{Dynamic Programming and Optimal Control}.\hskip 1em plus
	0.5em minus 0.4em\relax Athena Scientific, 2005, vol.~1.
	
	\bibitem{Abate10}
	A.~Abate, J.-P. Katoen, J.~Lygeros, and M.~Prandini, ``Approximate model
	checking of stochastic hybrid systems,'' \emph{European Journal of Control},
	vol.~16, no.~6, pp. 624--641, 2010.
	
	\bibitem{SoudjaniSiam13}
	S.~Soudjani and A.~Abate, ``Adaptive and sequential gridding procedures for the
	abstraction and verification of stochastic processes,'' \emph{{SIAM} Journal
		on Applied Dynamical Systems}, vol.~12, no.~2, pp. 921--956, 2013.
	
	\bibitem{Julius09}
	A.~A. Julius and G.~J. Pappas, ``Approximations of stochastic hybrid systems,''
	\emph{{IEEE} Transactions on Automatic Control}, vol.~54, no.~6, pp. 1193 --
	1203, 2009.
	
	\bibitem{Franzle11}
	M.~Fr{\"a}nzle, E.~M. Hahn, H.~Hermanns, N.~Wolovick, and L.~Zhang,
	``Measurability and safety verification for stochastic hybrid systems,'' in
	\emph{Hybrid Systems: Computation and Control}, 2011, pp. 43--52.
	
	\bibitem{Zamani14}
	M.~Zamani, I.~Tkachev, and A.~Abate, ``Bisimilar symbolic models for stochastic
	control systems without state-space discretization,'' in \emph{Hybrid
		Systems: Computation and Control}, 2014, pp. 41 -- 50.
	
	\bibitem{Lusena2001}
	C.~Lusena, J.~Goldsmith, and M.~Mundhenk, ``Nonapproximability results for
	partially observable {M}arkov decision processes,'' \emph{Journal of
		Artificial Intelligence Research}, vol.~14, pp. 83--103, 2001.
	
	\bibitem{shani13}
	G.~Shani, J.~Pineau, and R.~Kaplow, ``A survey of point-based {POMDP}
	solvers,'' \emph{Autonomous Agents and Multi-Agent Systems}, vol.~27, no.~1,
	pp. 1--51, 2013.
	
	\bibitem{sondik}
	E.~Sondik, ``The optimal control of partially observable {M}arkov processes,''
	Ph.D. dissertation, Stanford University, 1971.
	
	\bibitem{Brooks06}
	A.~Brooks, A.~Makarenko, S.~Williams, and H.~Durrant-Whyte, ``Parametric
	{POMDP}s for planning in continuous state spaces,'' \emph{Robotics and
		Autonomous Systems}, vol.~54, no.~11, pp. 887--897, 2006.
	
	\bibitem{Zhou2010}
	E.~Zhou, M.~Fu, and S.~Marcus, ``Solving continuous-state {POMDP}s via density
	projection,'' \emph{{IEEE} Transactions on Automatic Control}, vol.~55,
	no.~5, pp. 1101--1116, 2010.
	
	\bibitem{Porta06}
	J.~M. Porta, N.~Vlassis, M.~T. Spain, and P.~Poupart, ``Point-based value
	iteration for continuous {P}{O}{M}{D}{P}s,'' \emph{Journal of Machine
		Learning Research}, vol.~7, pp. 2329--2367, 2006.
	
	\bibitem{Brunskill2010}
	E.~Brunskill, L.~Kaelbling, T.~Lozano-Perez, and N.~Roy, ``Planning in
	partially-observable switching-mode continuous domains,'' \emph{Annals of
		Mathematics and Artificial Intelligence}, vol.~58, pp. 185--216, 2010.
	
	\bibitem{bertsekas}
	D.~P. Bertsekas and S.~E. Shreve, \emph{Stochastic Optimal Control: The
		Discrete-Time Case}.\hskip 1em plus 0.5em minus 0.4em\relax Athena
	Scientific, 1996.
	
	\bibitem{pineau06}
	J.~Pineau, G.~Gordon, and S.~Thrun, ``Anytime point-based approximations for
	large {POMDP}s,'' \emph{Journal of Artificial Intelligence Research},
	vol.~27, pp. 335--380, 2006.
	
	\bibitem{LesserArxiv_TAC14}
	K.~Lesser and M.~Oishi, ``Approximate verification of partially observable
	discrete time stochastic hybrid systems,'' 2014, preprint,
	\url{http://arxiv.org/abs/1410.8054}.
	
	\bibitem{Park91}
	J.~Park and I.~Sandberg, ``Universal approximation using radial-basis-function
	networks,'' \emph{Neural Computation}, vol.~3, no.~2, pp. 246 -- 257, 1991.
	
	\bibitem{Fornberg11}
	B.~Fornberg and N.~Flyer, ``The {G}ibbs phenomenon for radial basis
	functions,'' in \emph{Gibbs Phenomenon in Various Representations and
		Applications}, 2011, pp. 201--224.
	
	\bibitem{Zhang2010}
	K.~Zhang and J.~Kwok, ``Simplifying mixture models through function
	approximation,'' \emph{{IEEE} Transactions on Neural Networks}, vol.~21,
	no.~4, pp. 644--658, 2010.
	
	\bibitem{SARSOP}
	H.~Kurniawati, D.~Hsu, and W.~S. Lee, ``{SARSOP}: Efficient point-based {POMDP}
	planning by approximating optimally reachable belief spaces,'' in
	\emph{Robotics: Science and Systems}, 2008.
	
\end{thebibliography}

\end{document}